%% file: main.tex
\documentclass[aps,prb,onecolumn,superscriptaddress,nofootinbib,11pt]{revtex4-2}

\usepackage{amsmath,amssymb,bm}
\usepackage{graphicx}
\usepackage{booktabs}
\usepackage{siunitx}
\usepackage{xcolor}
\usepackage{tikz}
\usetikzlibrary{positioning,fit,backgrounds,arrows.meta}
\usepackage{pifont}
\usepackage{hyperref}
\providecommand{\PhonoFigureFont}{\sffamily}

\newcommand{\bq}{\mathbf{q}}
\newcommand{\br}{\mathbf{r}}
\newcommand{\kP}{\kappa_{\mathrm{P}}}
\newcommand{\kC}{\kappa_{\mathrm{C}}}
\begin{document}

\title{Phono3GPU: GPU-Accelerated Linear Tetrahedron Integration for Phonon Thermal Transport}

\author{Jinhyeong Jo}
\email{jjh1995@kist.re.kr}
\affiliation{Computational Science Research Center, Korea Institute of Science and Technology, Seoul, Republic of Korea}
\author{Sungwoo Kang}
\affiliation{Computational Science Research Center, Korea Institute of Science and Technology, Seoul, Republic of Korea}
\author{Byungju Lee}
\affiliation{Computational Science Research Center, Korea Institute of Science and Technology, Seoul, Republic of Korea}
\author{Jung-Hoon Lee}
\affiliation{Computational Science Research Center, Korea Institute of Science and Technology, Seoul, Republic of Korea}
\affiliation{KU-KIST Graduate School of Converging Science and Technology, Korea University, Seoul 02841, Republic of Korea}
\author{Seugchul Kim}
\email{sckim@kist.re.kr}
\affiliation{Computational Science Research Center, Korea Institute of Science and Technology, Seoul, Republic of Korea}

\begin{abstract}
Materials with high or low thermal conductivity are increasingly important for thermal management and energy
technologies. Machine-learned interatomic potentials have accelerated anharmonic force-constant calculations, shifting
the bottleneck toward phonon transport calculations for large materials datasets. Existing GPU solvers often rely on
smearing, whose width can require material-dependent choices, while the linear tetrahedron method used in phono3py
avoids this parameter but becomes costly for complex structures. Here we introduce Phono3GPU, which reformulates
three-phonon scattering and linear tetrahedron integration for massively parallel GPU execution. Across benchmark
crystals, Phono3GPU accelerates total calculations by up to 41 times relative to a CPU implementation, with larger
gains for complex unit cells, and by 2 to 6 times relative to a recent GPU plug-in for phono3py. We further demonstrate atom-resolved scattering analysis in defective silicon supercells
as an application enabled by the reduced computational cost. Phono3GPU thus lowers the cost of tetrahedron-based
thermal-conductivity calculations and of detailed investigations of structure-dependent phonon transport.
\end{abstract}

\maketitle

% =====================================================================
\section{Introduction}
\label{sec:intro}
% =====================================================================

Removing heat has become an increasingly important challenge in power electronics, highly integrated circuits, and solid-state lighting devices~\cite{Moore2014,Qian2021}. While materials with high thermal conductivity are essential for efficient heat dissipation and reliable device operation, materials with low thermal conductivity are equally important for thermoelectric energy conversion and thermal barrier coatings~\cite{Snyder2008,Padture2002}. Identifying materials with such contrasting thermal properties requires the efficient evaluation of thermal conductivity across a broad range of candidates. However, experimental characterization of the vast number of potential materials remains impractical, making lattice thermal conductivity calculations based on anharmonic lattice dynamics and the phonon Boltzmann transport equation (BTE) an important tool for materials screening.

Recent advances in universal machine-learned interatomic potentials (MLIPs) have substantially reduced the computational cost of obtaining the force constants required for these calculations~\cite{Deng2023,Batatia2023,Park2024}. With computational accelerators such as GPUs, the force evaluations for the displaced supercells have become far less costly~\cite{Pota2024}. These advances have expanded the accessible range of materials to complex crystals and structures containing defects or interfaces that were previously computationally prohibitive. However, while force-constant calculations have become considerably more efficient, the subsequent evaluation of phonon thermal transport remains computationally demanding. Widely used packages such as phono3py~\cite{Togo2015,Togo2023}, ShengBTE~\cite{Li2014}, and almaBTE~\cite{Carrete2017} primarily rely on CPU-based calculations. Consequently, phonon transport calculations are becoming an increasingly significant bottleneck as structural complexity and system size grow.

To address this bottleneck, several GPU-accelerated phonon transport programs have been developed. These implementations typically employ fixed or adaptive smearing to approximate energy conservation in phonon scattering processes. Although computationally efficient, smearing methods can be sensitive to the choice of broadening width, which may require additional convergence tests. This becomes particularly challenging in high-throughput materials screening, where numerous materials with diverse phonon properties must be evaluated systematically. In contrast, the linear tetrahedron method employed in phono3py treats energy conservation without introducing a smearing width, reducing the need for material-dependent parameter selection. However, its CPU-based implementation remains computationally expensive for large and complex structures, particularly those containing many atoms or exhibiting low symmetry. An efficient implementation of the tetrahedron method is therefore desirable for extending phonon thermal transport calculations to large-scale materials screening.

Here, we introduce Phono3GPU, a GPU-accelerated phonon thermal transport program that evaluates three-phonon and isotope scattering linewidths using the linear tetrahedron method. By reformulating the computationally demanding operations for efficient GPU parallelization, Phono3GPU calculates lattice thermal conductivity within the relaxation-time approximation (RTA), including the Wigner coherence contribution~\cite{Simoncelli2019,Simoncelli2022} (Fig.~\ref{fig:overview}). We validate its numerical accuracy against phono3py and systematically evaluate its computational performance (Fig.~\ref{fig:valid}). To demonstrate the practical benefits of this acceleration, we apply Phono3GPU to atom-resolved phonon scattering analysis in defective Si supercells (Fig.~\ref{fig:defect}) and extend the analysis to Si/diamond interfaces and diamond grain boundaries (Fig.~\ref{fig:iface}). These applications demonstrate the potential of Phono3GPU for efficient calculations of complex materials and detailed investigations of their phonon transport properties.

% =====================================================================
\section{Method}
\label{sec:gpu}
% =====================================================================
\input{fig_overview}

Existing GPU implementations of three-phonon scattering typically impose energy conservation using smearing methods~\cite{Barbalinardo2026,Zhang2021,Guo2026,Cepellotti2022}. These methods introduce a smearing width, chosen by the user or set adaptively, whereas the linear tetrahedron method introduces no such parameter. While this manuscript was in preparation, a third-party GPU plug-in for phono3py that includes the tetrahedron method was released~\cite{phono3pygpu}. It is compared with the present code in Fig.~\ref{fig:valid}c.
The present work recasts the calculation as operations on whole arrays, adds atom- and channel-resolved analyses of scattering, and applies them to cells of up to 312 atoms.

In the RTA, the computational cost is dominated by the evaluation of phonon linewidths. For each irreducible $\bq$ point, the linewidth is a sum over $\bq$-point triplets, and each triplet requires two quantities for all combinations of three phonon branches: the three-phonon interaction strength and the integration weight that imposes energy conservation. These quantities are independent. The interaction strength is obtained from the third-order force constants and phonon eigenvectors, whereas the integration weights depend only on the phonon frequencies. They are computed separately and then combined (Fig.~\ref{fig:overview}). phono3py evaluates these quantities using loops with data-dependent branches. To preserve its numerical conventions while enabling efficient GPU execution, we reformulated the three computational steps highlighted in Fig.~\ref{fig:overview} (Appendix~\ref{app:method}):

\begin{enumerate}
\item \emph{Fourier transform and mode projection as batched contractions.}
The Fourier transform of the third-order force constants and their projection onto phonon normal modes are expressed as batched matrix operations, allowing multiple $\bq$-point triplets to be processed simultaneously on the GPU.

\item \emph{Branch-free tetrahedron weights.}
The piecewise expressions of the linear tetrahedron method are evaluated over entire arrays, and masks select the appropriate region for each element instead of data-dependent branches. This ensures that the execution path does not depend on the input data.

\item \emph{Pruning.}
A tetrahedron contributes only when the energy lies between its lowest and highest vertex energies. Accordingly, only entries that can have non-zero integration weights are evaluated and stored, eliminating unnecessary operations and reducing the size of intermediate arrays.
\end{enumerate}

The interaction strengths and integration weights are combined through temperature-independent sums over phonon branches, after which the occupation factors are applied to obtain linewidths at all temperatures. Phonon frequencies and eigenvectors, velocity operators, isotope scattering, and the Wigner coherence term are also computed on the GPU. Input processing, grid construction, symmetry operations, and triplet generation remain on the CPU and use the phono3py Python API.

Within these steps, several operations are arranged so that less work is done for the same result. The two phase factors of the Fourier transform are multiplied before the sum over the cell images, which turns the double sum into a single matrix product. Band pairs that cannot fulfil energy conservation for any mode of the $\bq$ point are removed before the tetrahedron vertices are sorted. The isotope overlaps are formed only where the integration weight is non-zero. These arrangements change the results only at the level of rounding (Appendix~\ref{app:method}).

Phono3GPU is implemented in PyTorch~\cite{Paszke2019}, accepts force constants in phono3py format, and generates output files compatible with phono3py.

% =====================================================================
\section{Results}
\label{sec:results}
% =====================================================================
\begin{figure*}[tbh!]
\centering
\includegraphics[width=\textwidth]{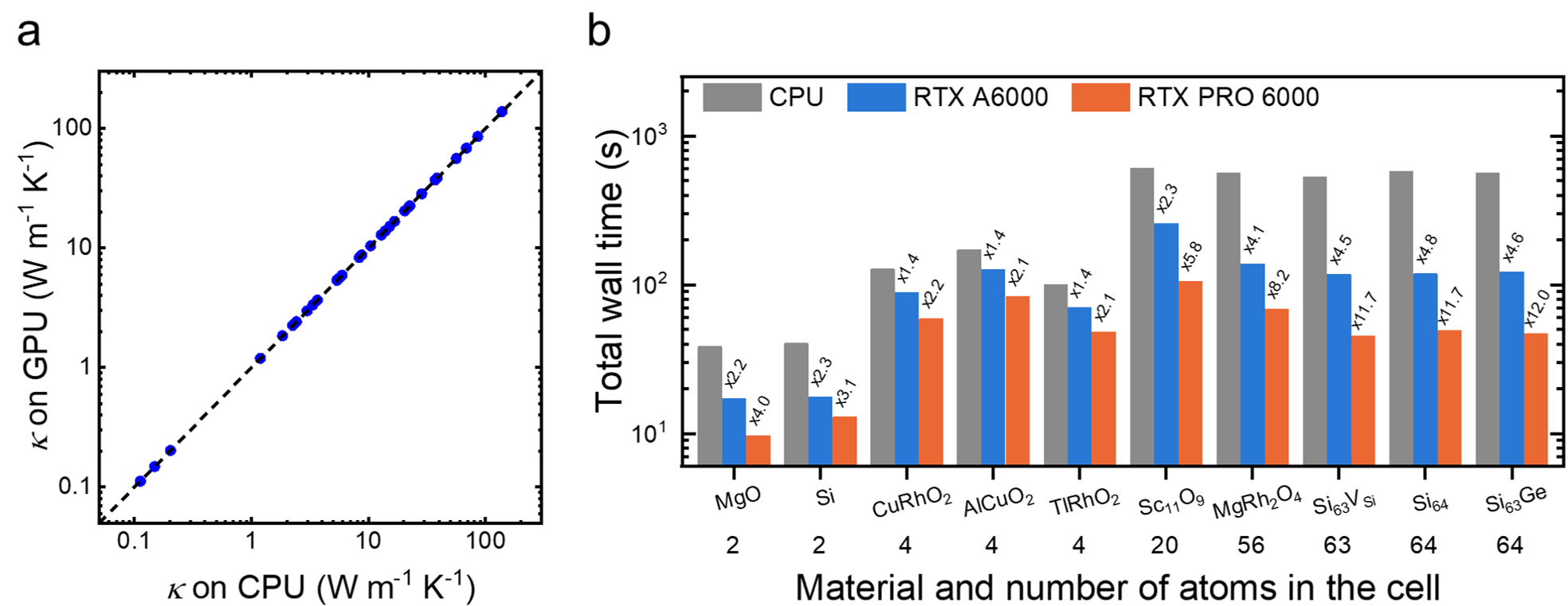}
\caption{Numerical validation and computational performance of Phono3GPU.
(a) Lattice thermal conductivities compared with phono3py~4.10 and phono3py-wte using identical force constants and meshes. The validation includes five dynamically stable crystals and TlRhO$_2$, which is included only as a numerical test.
(b) Total wall times on the CPU reference (32 threads), RTX A6000, and RTX PRO 6000 Blackwell. Numbers indicate speedups relative to the CPU, with atom counts shown beneath the material names.
(c) Total wall times of the GPU plug-in for phono3py~\cite{phono3pygpu} and of Phono3GPU on the RTX PRO 6000 Blackwell for the same calculations. Numbers indicate the ratio of the two times.}
\label{fig:valid}
\end{figure*}

An important requirement in reformulating phonon thermal transport calculations for GPU execution is to improve computational efficiency without compromising numerical accuracy. To verify this, we compared Phono3GPU with phono3py~4.10 using identical force constants and $\bq$-point meshes. The reference Wigner contribution was obtained using the phono3py-wte plug-in (Appendix~\ref{app:valid}). For the five dynamically stable benchmark crystals and the Si validation cells, the particle contribution ($\kP$) and Wigner coherence contribution ($\kC$) agree with the reference within normalized deviations of $3\times10^{-10}$ without isotope scattering. With natural isotope scattering, the maximum relative difference in total thermal conductivity is $3.4\times10^{-5}$ for the benchmark crystals and $5.2\times10^{-5}$ for the Si cells. The three-phonon linewidths also agree closely, with normalized deviations below $3\times10^{-8}$ even for the nearly degenerate modes of AlCuO$_2$. The dynamically unstable TlRhO$_2$ was evaluated separately as a numerical test (Appendix~\ref{app:valid}). These results demonstrate that Phono3GPU reproduces the reference calculations with high numerical accuracy despite the substantial restructuring of the computational operations (Fig.~\ref{fig:valid}a).

Having established numerical agreement, we evaluated the computational performance of Phono3GPU. We compared the total wall time measured from program launch to completion and found that Phono3GPU reduces the computational cost relative to the CPU implementation, with more pronounced benefits for larger and more complex structures. For calculations including natural isotope scattering, the RTX A6000 achieved speedups of 3.0 to 3.5 times for cells containing 2 to 4 atoms, 10 to 20 times for cells containing 20 to 56 atoms, and 17 to 20 times for Si supercells with 64 atomic sites. This trend became more pronounced on the RTX PRO 6000 Blackwell, where the corresponding speedups reached 2.7 to 4.6, 17 to 41, and 34 to 37 times, respectively (Fig.~\ref{fig:valid}b). The reference phono3py calculations used 32 threads on a 40-core CPU node, whereas Phono3GPU used a single GPU with four host CPU cores.

We also compared Phono3GPU with the GPU plug-in for phono3py~\cite{phono3pygpu}, which evaluates the same tetrahedron integration on the GPU and is therefore the most direct reference among the GPU codes. With identical inputs, meshes and isotope settings on the RTX PRO 6000 Blackwell, the total wall time of Phono3GPU is 4.2 to 6.5 times shorter for the cells containing 2 to 4 atoms and 2.0 to 3.5 times shorter for the cells containing 20 to 64 atoms (Fig.~\ref{fig:valid}c). The corresponding values on the RTX A6000 are 5.3 to 7.9 and 2.2 to 3.7 (Appendix~\ref{app:perf}).

The computational gains demonstrated for the 64-site Si supercells provide a basis for extending the analysis to larger and more complex atomic structures. Building on these results, we applied Phono3GPU to 128-site Si supercells containing two point defects and to interface structures containing up to 312 atoms. The thermal conductivity calculations for the 128-site defect supercells were completed within 40 to 74 minutes on an RTX A6000 (Appendix~\ref{app:perf}). These applications demonstrate the practical use of Phono3GPU for investigating phonon scattering and thermal transport in complex defect and interface structures.

% =====================================================================

\label{sec:defect}

\begin{figure*}[tbh!]
\centering
\includegraphics[width=\textwidth]{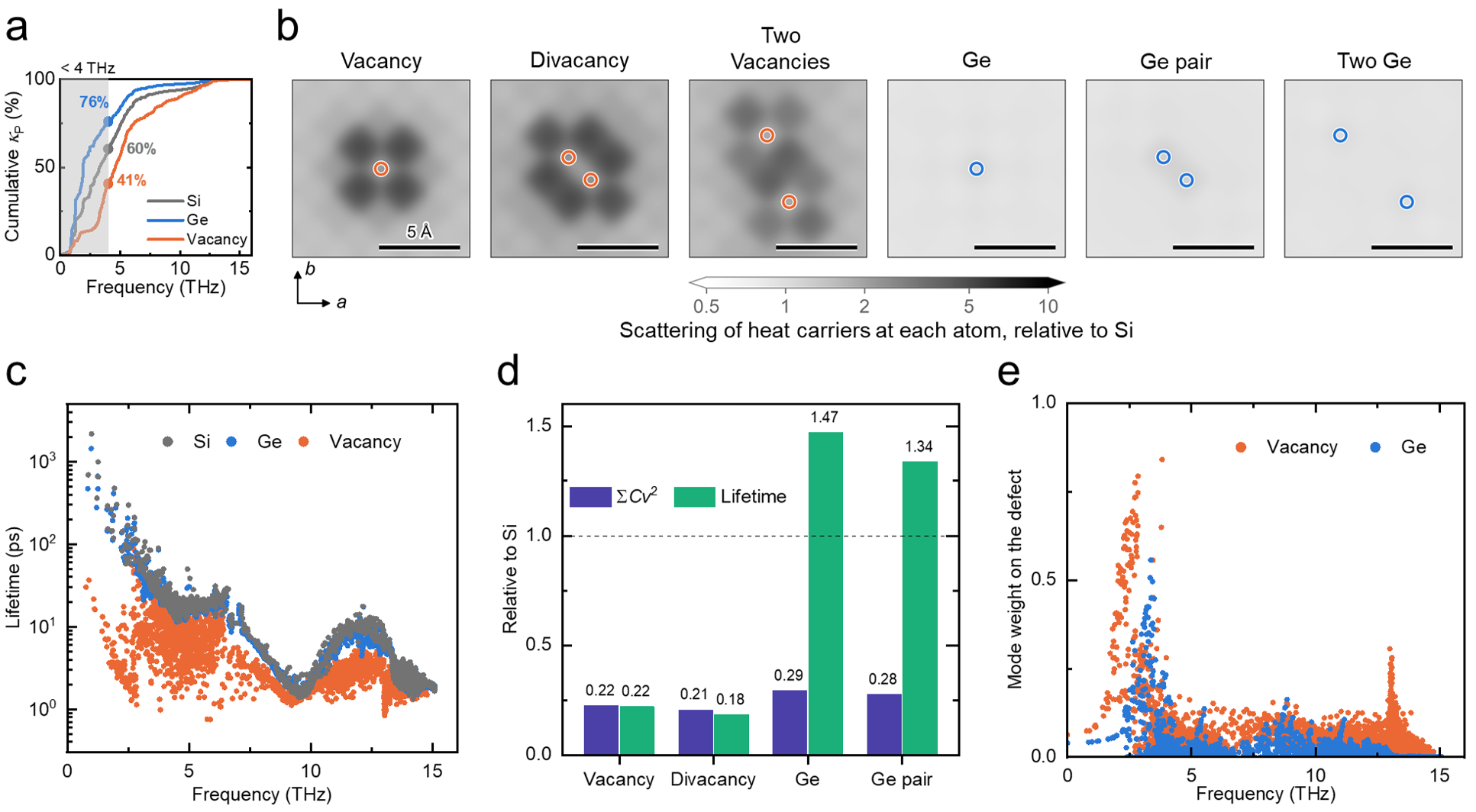}
\caption{
Phonon transport and atom-resolved scattering in defective Si calculated using the Prophet potential at 300~K with natural isotopes.
(a) Cumulative particle thermal conductivity $\kP$ versus frequency for perfect Si, one vacancy, and one Ge substitution.
(b) Atom-resolved contributions $N\Gamma_i$ to heat-carrier scattering relative to perfect Si.
The six configurations include single vacancies and Ge substitutions in 64-site cells, adjacent defect pairs in 64-site cells, and separated defect pairs in 128-site cells.
Orange and blue circles indicate vacancies and Ge atoms, respectively.
The maps are projected along $c$ by averaging the contributions of atoms within each column.
(c) Mode-resolved phonon lifetimes including three-phonon and isotope scattering.
(d) Decomposition of $\kP$ relative to perfect Si into $\sum Cv^2$ and the $Cv^2$-weighted average lifetime.
(e) Vibrational weights of phonon modes on the Ge atom or the four atoms neighbouring a vacancy.}
\label{fig:defect}
\end{figure*}

We first demonstrate the application of Phono3GPU to atom-resolved phonon scattering analysis in defective Si supercells. Spatially resolved descriptions of heat transport around defects are available from other approaches. Gautam \textit{et al.} projected the thermal conductivity itself onto atomic sites within a harmonic theory for vacancies, SiGe alloys and grain boundaries~\cite{Gautam2024}, and Yokoi \textit{et al.} resolved the thermal conductivity by atom around point defects in PbTe using molecular dynamics with a machine-learned potential, which incorporates anharmonicity within classical molecular dynamics~\cite{Yokoi2025}. The quantity resolved here is different. Within the BTE, the three-phonon interaction amplitude can be decomposed into contributions from individual atoms, allowing the phonon linewidth to be partitioned into atomic contributions whose sum exactly recovers the total linewidth (Appendix~\ref{si:atom}). The isotope contribution can be partitioned in the same manner. From this decomposition, we evaluate $\Gamma_i$, which represents the atomic contribution to the scattering of heat-carrying phonons, averaged over phonon modes with the transport weight $Cv^2$. This partition retains interference between atomic contributions rather than defining an independent local scattering rate. The same computational procedure also resolves the contributions from different scattering channels and the atomic weights of individual phonon modes. For the 64-site defect supercells, the atom-resolved analysis requires 150 to 165 seconds and approximately 19~GB of GPU memory, corresponding to about 1.5 times the cost of a standard thermal conductivity calculation. The decomposition therefore comes at the cost of roughly one additional conductivity calculation, which makes it practical to apply to every defect configuration considered below.

To examine how different point defects influence phonon transport, we considered three configurations for each defect type. These include a 64-site Si supercell containing one vacancy or one Ge substitution at a concentration of 1.6\%, a 64-site cell containing a nearest-neighbour defect pair at 3.1\%, and a 128-site cell containing two separated defects at 1.6\%. The force constants were obtained using the universal machine-learned interatomic potential Prophet-OAME-MBD~\cite{Prophet}, which has been evaluated in the Matbench Discovery benchmark~\cite{Riebesell2025}, whose thermal-conductivity metric follows Ref.~\cite{Pota2024}. The lattice parameters were fixed at the equilibrium values predicted by the potential, while the atomic positions were relaxed. Thermal conductivities were calculated at 300~K with natural isotope scattering using a $6\times6\times6$ mesh for the 64-site cells and a $6\times6\times3$ mesh for the 128-site cells. The third-order force constants were evaluated without a distance cutoff for the 64-site cells, while a cutoff of 6.3~\AA\ was applied to the 128-site cells. Because the defects are periodically repeated, these calculations describe ordered defect arrangements and do not include a separate elastic scattering contribution from randomly distributed Ge substitutions.

The thermal conductivity of perfect Si was calculated to be 124.2~W\,m$^{-1}$K$^{-1}$, decreasing to 8.2~W\,m$^{-1}$K$^{-1}$ with a single vacancy and 58.1~W\,m$^{-1}$K$^{-1}$ with one Ge substitution. At the same defect concentration of 1.6\%, a vacancy therefore reduces the total thermal conductivity by approximately a factor of 15, compared with about a factor of two for Ge substitution. This contrast persists across the other defect configurations. The three vacancy-containing structures yield conductivities from 6.0 to 8.2~W\,m$^{-1}$K$^{-1}$, whereas the Ge-substituted structures range from 48.3 to 58.1~W\,m$^{-1}$K$^{-1}$ (Table~\ref{tab:defectcells}). To understand why these reductions differ, we examine the phonon transport and scattering characteristics of the single-defect structures, using the remaining configurations to assess whether the same trends persist.

We first examined the frequency distribution of phonons contributing to heat transport. In perfect Si, modes below 3.3~THz account for half of the particle conductivity $\kP$. This frequency decreases to 1.9~THz upon Ge substitution but increases to 4.6~THz in the presence of a vacancy (Fig.~\ref{fig:defect}a). These changes indicate that the two defects redistribute the contributions of phonon modes to thermal conductivity in different ways.

To examine how these changes relate to local scattering, we evaluated the spatial distribution of the atom-resolved contribution $\Gamma_i$ (Fig.~\ref{fig:defect}b). In the vacancy-containing structures, the largest contributions occur on atoms neighbouring the vacancies and reach 9.8 to 12.8 times the corresponding value in perfect Si. In contrast, the maximum contributions in the Ge-substituted structures are only 1.35 to 1.62 times that of Si. For the single-vacancy cell, the average scattering contribution over the cell is approximately twice that of perfect Si. For the single-Ge cell, the corresponding average is about 0.9 times the Si value despite the local increase around Ge. These observations indicate that vacancies produce a pronounced enhancement of scattering contributions near the defects, whereas Ge substitution changes the local distribution without increasing the cell-averaged contribution.

The mode-resolved phonon lifetimes support this distinction (Fig.~\ref{fig:defect}c). In the single-vacancy structure, the lifetimes of modes below 3~THz are approximately 15 to 40 times shorter than those in perfect Si. In contrast, the lifetimes in the Ge-substituted structure remain close to those of Si at comparable frequencies. The combination of atom-resolved scattering contributions and phonon lifetimes therefore indicates that enhanced scattering plays an important role in the conductivity reduction caused by vacancies, while the reduction associated with Ge substitution requires a different explanation.

To distinguish the factors responsible for this reduction, we separated the particle conductivity $\kP$ into two factors (Fig.~\ref{fig:defect}d). The first is $(VN_q)^{-1}\sum_\lambda C_\lambda v_\lambda^2$, denoted by $\sum Cv^2$, which represents the contribution of phonon heat capacities and group velocities. Here, $V$ is the cell volume, $N_q$ is the number of $\bq$ points, and $v_\lambda^2$ is the squared group velocity averaged over the three Cartesian directions. The second factor is the $Cv^2$-weighted average phonon lifetime. For a single vacancy, both factors decrease to approximately 0.22 of their values in perfect Si. In contrast, Ge substitution reduces $\sum Cv^2$ to 0.29, while the weighted lifetime factor increases to 1.47. The same distinction persists across the other defect configurations. For the vacancy-containing structures, both factors range from 0.18 to 0.22, whereas the Ge-substituted structures exhibit $\sum Cv^2$ ratios of 0.27 to 0.29 and weighted lifetime ratios of 1.34 to 1.47 (Table~\ref{tab:defectcells}).

The increase in the weighted lifetime factor for Ge does not indicate a general increase in individual phonon lifetimes, which remain close to those of perfect Si at comparable frequencies (Fig.~\ref{fig:defect}c). Instead, the contribution to $\sum Cv^2$ below 2~THz remains nearly unchanged at 1.02 times that of Si, while contributions in the frequency intervals above 2~THz decrease to 0.25 to 0.34 times their original values. This redistribution increases the relative weight of longer-lived low-frequency modes, consistent with the cumulative conductivity in Fig.~\ref{fig:defect}a.

These results reveal two distinct mechanisms of conductivity reduction. Vacancies suppress both the $\sum Cv^2$ factor and phonon lifetimes, reducing $\kP$ to approximately 0.049 of its value in perfect Si. In contrast, Ge substitution primarily reduces $\sum Cv^2$, while the weighted lifetime factor remains above unity, resulting in a $\kP$ ratio of approximately 0.43. The Wigner coherence contribution $\kC$ increases in both cases, from 0.15~W\,m$^{-1}$K$^{-1}$ in perfect Si to 2.08 and 4.49~W\,m$^{-1}$K$^{-1}$ for the vacancy and Ge substitution, respectively. However, this increase does not compensate for the reduction in $\kP$, leading to the larger decrease in total thermal conductivity for vacancies than for Ge substitutions (Table~\ref{tab:defectcells}).

These two mechanisms can be related to the earlier atom-resolved studies. The harmonic site projection of the conductivity found that the under-coordinated atoms next to a Si vacancy carry about 70\% less conductivity than atoms of the bulk crystal~\cite{Gautam2024}, and the molecular dynamics study of point defects in PbTe concluded that the change of the harmonic force field around the defects plays a central role in the reduction~\cite{Yokoi2025}. Those studies resolve the conductivity itself and do not separate a harmonic transport factor from the phonon lifetimes. The present decomposition makes this separation. For Ge the loss of conductivity comes from $\sum Cv^2$ while the scattering of heat carriers does not increase, and for the vacancy three-phonon scattering increases on the neighbouring atoms and contributes a factor of the same size as the harmonic one.

Finally, we examined the spatial distribution of phonon vibrations around the defects (Fig.~\ref{fig:defect}e). In the Ge-substituted structure, selected modes between 2.6 and 3.5~THz carry approximately 38\% of their vibrational weight on the Ge atom, compared with 1.6\% for a uniform distribution. Similarly, selected modes between 1.8 and 3.8~THz in the vacancy structure carry approximately 50\% of their weight on the four neighbouring atoms, compared with approximately 6\% for a uniform distribution. These modes exhibit low participation ratios of approximately 0.10 to 0.20, consistent with vibrational weight concentrated around the defects. Nevertheless, their contribution to $\kP$ is less than 1\%. Additional atom-resolved analyses, channel-resolved scattering contributions, and the dependence of thermal conductivity on defect separation are provided in Appendix~\ref{si:pairs}.

Together, these results demonstrate how atom-resolved scattering contributions, phonon lifetimes, and the $\sum Cv^2$ decomposition can be combined to distinguish the thermal transport characteristics of different defects. Phono3GPU makes such detailed analyses practical for complex defect supercells, illustrating its potential for investigating phonon transport in larger atomic structures.

% =====================================================================
\label{sec:interface}

\begin{figure*}[tbh!]
\centering
\includegraphics[width=\textwidth]{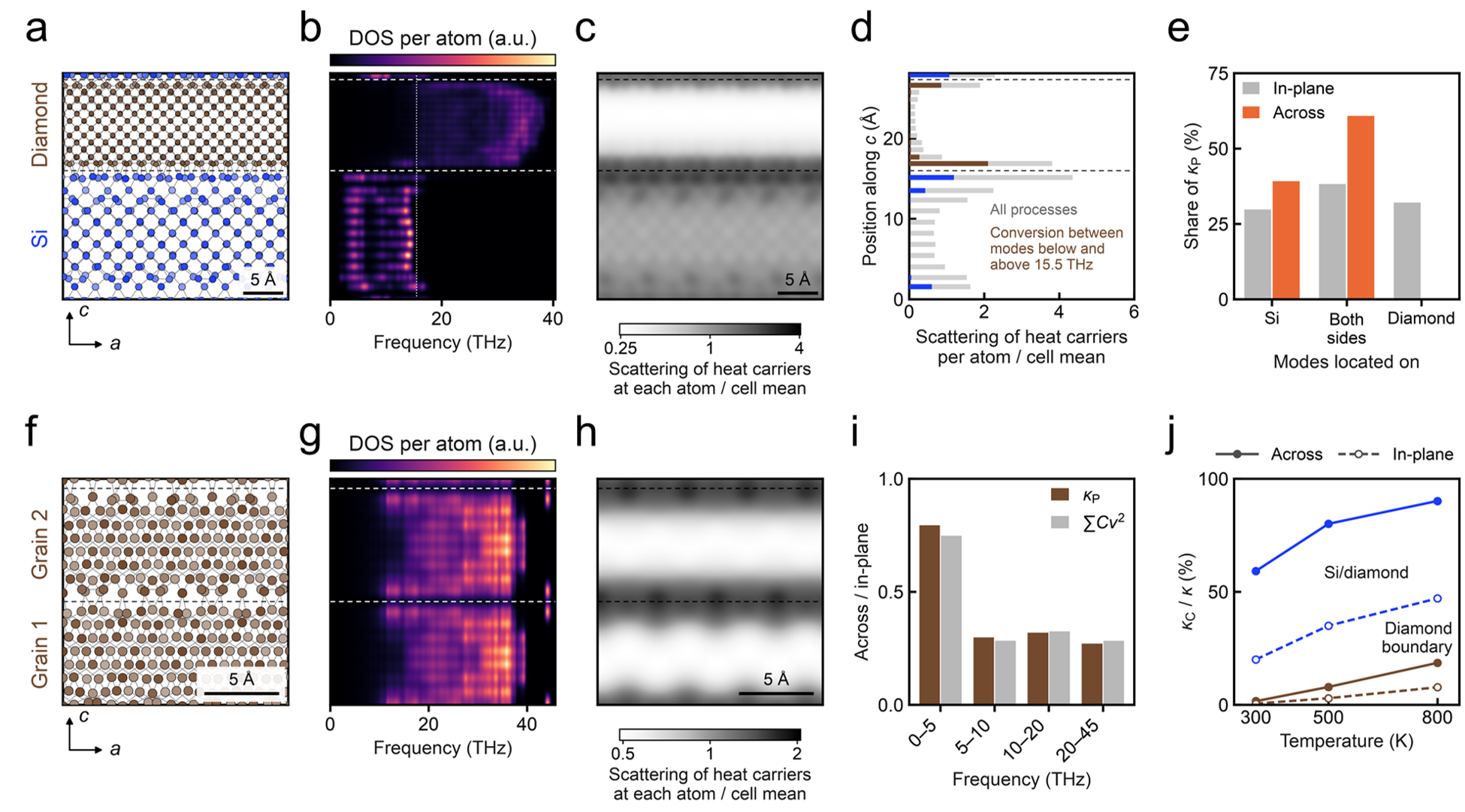}
\caption{
Phonon transport in Si/diamond interfaces and diamond grain boundaries calculated using the Prophet potential within RTA and Wigner theory at 300~K with natural isotopes, unless otherwise stated. (a) to (e) Si/diamond (001) superlattice containing 312 atoms. (a) Atomic structure viewed along the $b$ direction. (b) Layer-resolved vibrational density of states. The dotted line marks the upper limit of the Si phonon spectrum at 15.5~THz. (c) Spatial distribution of atom-resolved scattering contributions $\Gamma_i$, normalized by the cell average. (d) Layer-resolved scattering contributions and frequency-conversion channels across 15.5~THz. (e) Fractions of in-plane and cross-plane $\kP$ carried by modes localized in Si, diamond, or distributed across both materials. Localization is defined by more than 90\% of the vibrational weight. (f) to (i) Diamond (001) $\Sigma5$ twist grain boundary containing 160 atoms. (f) Atomic structure. (g) Layer-resolved vibrational density of states. (h) Spatial distribution of atom-resolved scattering contributions. (i) Cross-boundary to in-plane ratios of $\kP$ and $\sum Cv^2$ in different frequency ranges. (j) Temperature dependence of the Wigner coherence fraction $\kC/\kappa$ for both structures.}
\label{fig:iface}
\end{figure*}

We next extended the atom-resolved scattering analysis of defective Si to larger and more complex interface structures. Interfacial thermal transport has been investigated using the phonon BTE, molecular dynamics, Green's function approaches and site-projected thermal conductivities~\cite{Garg2013,Khosravian2013,Gordiz2016,Guo2021,Gautam2024}. Here, we use the computational efficiency of Phono3GPU to complement these approaches with one specific quantity, the three-phonon linewidth resolved by atom and by scattering channel, examined together with the spatial distribution of the phonon modes. We considered two systems with distinct interfacial characteristics: a Si/diamond interface between materials with different vibrational spectra and a diamond grain boundary between regions of the same material (Fig.~\ref{fig:iface}).

Using the Prophet-OAME-MBD potential, we constructed a 312-atom Si/diamond (001) superlattice and a 160-atom diamond (001) $\Sigma5$ twist grain boundary. The Si/diamond superlattice contains 12 atomic layers of each material, with a lateral match between $2\times2$ Si and $3\times3$ diamond cells that imposes approximately 1.7\% compressive strain on Si. Periodic boundary conditions produce two interfaces, one of which reconstructs during structural relaxation along unstable zone-centre phonon modes. The reconstructed interface contains three- and five-coordinated atoms, whereas the other interface consists entirely of four-coordinated atoms. The diamond grain boundary contains 16 three-coordinated atoms, with two periodically repeated boundaries separated by 0.75~nm. Both structures were relaxed in their lattice parameters and atomic positions and were confirmed to be dynamically stable.

With a 6.3~\AA\ cutoff, the third-order force constants required 321,664 displaced supercells for Si/diamond and 40,384 for the diamond grain boundary. For the 312-atom structure, the atom-resolved and channel-resolved analysis of 936 phonon branches was completed in approximately 4.2 GPU-hours. The calculation used 26 irreducible $\bq$ points from a $6\times6\times2$ mesh. All conductivities discussed below were calculated within the RTA, including the Wigner coherence contribution. For reference, the same potential gives a bulk diamond conductivity of 1755~W\,m$^{-1}$K$^{-1}$ within the RTA and 2241~W\,m$^{-1}$K$^{-1}$ from the direct solution of the linearised BTE at 300~K with natural isotopes. The RTA therefore underestimates the bulk diamond conductivity by approximately 22\%, suggesting that a similar limitation may affect the interface calculations.

In addition to resolving scattering by atom, we separated the three-phonon linewidths into decay and coalescence processes and further classified the channels according to the frequencies of the participating phonons. This decomposition reproduces the total linewidth with a relative difference of approximately $10^{-15}$ and allows the spatial distribution of different scattering channels to be examined.

For the Si/diamond superlattice, the thermal conductivity at 300~K was 13.5~W\,m$^{-1}$K$^{-1}$ parallel to the interfaces and only 0.40~W\,m$^{-1}$K$^{-1}$ across them. To understand this strong anisotropy, we first examined the vibrational spectra and spatial distributions of the phonon modes. The spectra of Si and diamond overlap only below approximately 15.5~THz, corresponding to the upper limit of the Si phonon spectrum (Fig.~\ref{fig:iface}a,b). Approximately 61\% of the modes are predominantly confined to diamond. These modes contribute 32\% of the in-plane particle conductivity $\kP$ but negligibly to the cross-plane contribution. In contrast, modes extending across both materials account for only 21\% of all modes but carry 61\% of the cross-plane $\kP$ (Fig.~\ref{fig:iface}e). This indicates that phonon modes distributed across both materials play an important role in heat transport perpendicular to the interfaces.

The atom-resolved analysis further reveals that three-phonon scattering is strongly concentrated near the interfaces (Fig.~\ref{fig:iface}c). The atomic layers adjacent to the two interfaces contain only 19\% of the atoms but account for 58\% of the total scattering contribution from heat-carrying phonons. Because the in-plane conductivity is much larger than the cross-plane conductivity, this transport-weighted map primarily reflects the scattering of modes carrying heat parallel to the interfaces.

To examine the scattering processes associated with the spectral mismatch, we classified three-phonon channels using 15.5~THz as the frequency boundary (Fig.~\ref{fig:iface}d). Processes involving the decay of a high-frequency phonon into two lower-frequency phonons, or the reverse coalescence process, account for approximately 30\% of the total scattering contribution from heat-carrying phonons. Notably, 85\% of these frequency-conversion contributions are concentrated in the interface layers. Such processes account for 44\% of the scattering contribution within the interface layers, compared with only 11\% in the remaining regions. The reconstructed interface also exhibits approximately 1.8 times the scattering contribution of the fully coordinated interface. These results show that phonon frequency-conversion processes are concentrated near the Si/diamond interfaces and are sensitive to their local atomic structures.

The Wigner coherence contribution is also particularly important for cross-plane heat transport in this superlattice (Fig.~\ref{fig:iface}j). Its share of the cross-plane conductivity increases from 59\% at 300~K to 80\% at 500~K and 90\% at 800~K. Consequently, although the particle contribution decreases with increasing temperature, the total cross-plane conductivity rises from 0.40 to 0.59~W\,m$^{-1}$K$^{-1}$ between 300 and 800~K. At 800~K, neglecting the Wigner contribution would underestimate the cross-plane conductivity by approximately a factor of ten. Thus, the thermal transport characteristics of the Si/diamond superlattice involve both the spatial distribution of phonon modes and interface-localized anharmonic scattering, while the coherence contribution becomes dominant in the cross-plane direction.

In the Si/diamond superlattice, the vibrational mismatch between the two materials and the structural changes at the interfaces occur together. To examine a boundary without such a mismatch between the bulk materials, we performed the same analysis for the diamond grain boundary, where both sides have identical atomic masses and bulk phonon spectra (Fig.~\ref{fig:iface}f,g). At 300~K, the conductivity was 173~W\,m$^{-1}$K$^{-1}$ parallel to the boundary and 75~W\,m$^{-1}$K$^{-1}$ across it. The ratio between these directions remained approximately 0.43 from 300 to 800~K.

Unlike the pronounced directional reduction in conductivity, the $Cv^2$-weighted average phonon lifetime was not shorter across the grain boundary. It was approximately 29~ps in the cross-boundary direction compared with 22~ps in the parallel direction. To identify the origin of this anisotropy, we compared the frequency-dependent directional ratios of $\kP$ and $\sum Cv^2$ (Fig.~\ref{fig:iface}i). Below 5~THz, the cross-boundary $\kP$ was approximately 79\% of its in-plane value, decreasing to 30\% between 5 and 10~THz. Across the frequency intervals, the directional ratios of $\sum Cv^2$ closely followed those of $\kP$, with differences smaller than 0.04. The corresponding weighted lifetime ratios remained between 0.96 and 1.06. These results indicate that the reduction in cross-boundary conductivity is more closely associated with the directional changes in $\sum Cv^2$ than with additional shortening of phonon lifetimes.

Nevertheless, the grain boundary exhibits enhanced local scattering. The atom-resolved contribution from boundary atoms is approximately eight times that of atoms in the interior (Fig.~\ref{fig:iface}h). For modes between 10 and 20~THz, the boundary layers contain approximately 25\% of the atoms but account for 62\% of their scattering contribution. Thus, although scattering is locally enhanced near the grain boundary, the overall anisotropy of particle thermal conductivity is primarily associated with the directional difference in the harmonic transport factor. A harmonic site projection likewise found a reduced local conductivity at the boundary atoms of a Si twin boundary~\cite{Gautam2024}. Here the anharmonic scattering is resolved separately, and it is locally enhanced without being the origin of the anisotropy. The coherence contribution also increases with temperature, accounting for 1.6\% of the cross-boundary conductivity at 300~K and 18\% at 800~K, but remains less dominant than in the Si/diamond superlattice (Fig.~\ref{fig:iface}j).

The two interface systems therefore exhibit distinct thermal transport characteristics. In Si/diamond, the spatial separation of phonon modes and interface-localized frequency-conversion processes accompany a dominant coherence contribution to cross-plane transport. In the diamond grain boundary, local scattering is enhanced, but the conductivity anisotropy is more closely related to the directional difference in $\sum Cv^2$. These applications demonstrate that GPU acceleration with Phono3GPU makes detailed mode-resolved and atom-resolved analyses more practical for complex interface structures containing hundreds of atoms, reducing the computational burden of investigating their thermal transport properties.

% =====================================================================
\section{Conclusions}
% =====================================================================

We developed Phono3GPU by reformulating linear tetrahedron integration for three-phonon and isotope scattering to enable efficient GPU execution. Phono3GPU reproduces the results of phono3py with high numerical accuracy without introducing a smearing parameter, while reducing the total wall time by up to a factor of 41. The computational gains become more pronounced for larger and more complex structures. Using this efficiency, we distinguished the factors responsible for thermal conductivity reduction by vacancies and Ge substitutions in Si. We further examined the spatial distribution of phonon modes and atom-resolved and channel-resolved scattering contributions in Si/diamond interfaces and diamond grain boundaries. These applications demonstrate how GPU acceleration reduces the computational burden of detailed phonon transport analysis in complex structures containing hundreds of atoms.

More broadly, advances in computational methods, from efficient force-constant evaluation using machine-learned interatomic potentials to GPU-accelerated phonon transport calculations, are expanding the range of atomic structures accessible to detailed phonon analysis. One feature of this approach is that the phonon occupations and heat capacities follow Bose--Einstein statistics. The three-phonon scattering at defects and interfaces can therefore be analysed in the same way at low temperatures, where molecular dynamics with classical nuclei requires quantum corrections. At low temperatures three-phonon scattering becomes weaker and isotope and boundary scattering become relatively more important, so such an analysis has to be considered together with these processes. The present work does not include a direct comparison with other methods at low temperatures, which remains to be examined.

Such low-temperature phonon analysis may benefit materials research beyond high-throughput screening of materials with high or low thermal conductivity, in areas where phonon lifetimes and anharmonic scattering are important. Quantum acoustic devices are one example, in which phonon lifetimes limit the device performance and defects and interfaces can act as sources of loss. By enabling efficient analysis of which atoms and which scattering channels shorten the phonon lifetimes in larger systems containing defects and interfaces, such computational approaches may provide deeper insights into phonon dynamics and support materials design for diverse applications.

\section*{Data and code availability}

The Phono3GPU source code is available at \url{https://github.com/jjh-springbank/phono3gpu}. The data that support the findings of this study are available from the corresponding author upon reasonable request.

\acknowledgments This work was supported by the KIST Institutional Program (26E0223 and 26E0440) and the KU-KIST Graduate School of Converging Science and Technology (KU-KIST school program).

\appendix

\section{Formulation and implementation}\label{app:method}

We use $\lambda=(\bq,j)$ for a phonon mode, $p=1,\ldots,n_p$ and $s=1,\ldots,n_s$ for atoms in the
primitive cell and supercell, and $c=1,\ldots,n_c$ for lattice translations, with $n_c=n_s/n_p$ and $n_b=3n_p$
phonon bands. Greek indices denote Cartesian components, and superscript $b$ identifies one of $B$ triplets in a
GPU batch. Unless otherwise stated, numerical conventions follow phono3py. Only operations reformulated for the
GPU are described in detail.

\subsection{Quantities that are computed}

The three-phonon linewidth is~\cite{Togo2015}
\begin{multline}
\Gamma_{\lambda}(T)=\frac{18\pi}{\hbar^2}\sum_{\lambda'\lambda''}
\bigl|\Phi_{\lambda\lambda'\lambda''}\bigr|^2
\Bigl\{(n_{\lambda'}+n_{\lambda''}+1)\,\delta(\omega_\lambda-\omega_{\lambda'}-\omega_{\lambda''})\\
+(n_{\lambda'}-n_{\lambda''})\bigl[\delta(\omega_\lambda+\omega_{\lambda'}-\omega_{\lambda''})
-\delta(\omega_\lambda-\omega_{\lambda'}+\omega_{\lambda''})\bigr]\Bigr\},
\label{eq:gamma}
\end{multline}
where $n_\lambda$ is the Bose--Einstein occupation and the sum runs over the triplets with
$\bq+\bq'+\bq''$ equal to a reciprocal lattice vector. The interaction strength is
\begin{equation}
\Phi_{\lambda\lambda'\lambda''}=\frac{1}{\sqrt{N_q}}\,\frac{1}{3!}
\Bigl(\frac{\hbar^3}{8\,\omega_\lambda\omega_{\lambda'}\omega_{\lambda''}}\Bigr)^{1/2}V_{\lambda\lambda'\lambda''},
\label{eq:phinorm}
\end{equation}
where $V_{\lambda\lambda'\lambda''}$ is the projected third-order coupling defined in
Eqs.~\eqref{eq:r2r} and~\eqref{eq:proj}. Following phono3py, the triplet conserves crystal momentum and the
projection uses eigenvectors without complex conjugation. By time-reversal symmetry, this is equivalent to the
convention written with $\Phi_{-\lambda\lambda'\lambda''}$~\cite{Togo2015}. Energy conservation is treated by linear tetrahedron
integration~\cite{Lambin1984,Blochl1994}. Isotope scattering is given in Eq.~\eqref{eq:iso}. The Wigner
conductivity~\cite{Simoncelli2019} is
\begin{multline}
\kappa^{\alpha\beta}=\frac{1}{VN_q}\sum_{\bq}\sum_{jj'}
\frac{\omega_{\bq j}+\omega_{\bq j'}}{4}
\Bigl(\frac{C_{\bq j}}{\omega_{\bq j}}+\frac{C_{\bq j'}}{\omega_{\bq j'}}\Bigr)
\mathrm{Re}\bigl[v^\alpha_{\bq,jj'}v^\beta_{\bq,j'j}\bigr]\\
\times\frac{\Gamma_{\bq j}+\Gamma_{\bq j'}}
{(\omega_{\bq j}-\omega_{\bq j'})^2+(\Gamma_{\bq j}+\Gamma_{\bq j'})^2},
\label{eq:wigner}
\end{multline}
Here $V$ without mode indices is the primitive-cell volume, $C$ is the mode heat capacity, and $v_{\bq,jj'}$
is a velocity-operator matrix element. The total linewidth $\Gamma$ includes three-phonon, isotope and optional
boundary terms. With the convention of Eq.~\eqref{eq:gamma}, $\tau=1/(2\Gamma)$.

\subsection{Harmonic phonons, velocity operators and polar crystals}
\label{sec:nac}

The dynamical matrix is built for a batch of $\bq$ points at once,
\begin{equation}
D^{b}_{p\alpha,p'\beta}=\frac{1}{\sqrt{m_pm_{p'}}}\sum_{s\in p'}
\Phi^{(2)}_{p\alpha,s\beta}\,\phi_{\bq^{b}}(p,s),\qquad
\phi_{\bq}(p,s)=\frac{1}{M_{ps}}\sum_{m=1}^{M_{ps}}e^{2\pi i\,\bq\cdot\br^{(m)}_{ps}},
\label{eq:dynmat}
\end{equation}
Here $s\in p'$ denotes supercell images of $p'$, and $\br^{(m)}_{ps}$ are the $M_{ps}$ shortest vectors from $p$
to $s$. The matrices are diagonalised in a batched call. The velocity operator
$\partial\sqrt{D(\bq)}/\partial\bq$ is obtained by central differences with
$|\Delta\bq|=5\times10^{-6}$. The three acoustic modes at $\Gamma$ are excluded, as in phono3py. The
symmetry-star sum is evaluated by one contraction with the precomputed tensor
$\sum_{R\in\mathcal S(\bq)}R\otimes R$ for each $\bq$.

For polar crystals, the Gonze and Lee correction~\cite{Gonze1997} follows phonopy~\cite{Togo2023}.
Phonopy prepares the short-range force constants, on-site term, reciprocal vectors and Ewald parameter once on the
host. The $\bq$-independent phase factors $e^{2\pi i\mathbf G\cdot(\br_p-\br_{p'})}$ are cached on the GPU,
allowing the reciprocal-space sum for a batch of $\bq$ points to be evaluated as a tensor contraction. The
singular contribution with $\bq+\mathbf G=0$ is omitted. Near $\Gamma$, finite differences for the velocity
operator use the directional limits at $\bq\pm\Delta\bq$.

\subsection{Third-order force constants in reciprocal space}
\label{sec:fc3rec}

For a triplet $(\bq_0,\bq_1,\bq_2)$, phono3py evaluates
\begin{equation}
\Phi_{p_0p_1p_2}(\bq_0,\bq_1,\bq_2)=e^{2\pi i(\bq_0+\bq_1+\bq_2)\cdot\br_{p_0}}
\sum_{s_1\in p_1}\sum_{s_2\in p_2}\Phi^{(3)}_{p_0s_1s_2}\,\phi_{\bq_1}(p_0,s_1)\,\phi_{\bq_2}(p_0,s_2)
\label{eq:r2r}
\end{equation}
(Cartesian indices suppressed) in loops over the atoms and their images. We permute the supercell atoms so that
$s\mapsto(p,c)$, reshape $\Phi^{(3)}_{p_0s_1s_2}$ into $W_{p_0,p_1,c_1,p_2,c_2,x}$, where $x$ runs over the 27
Cartesian triples, and write Eq.~\eqref{eq:r2r} as two batched contractions,
\begin{align}
X^{b}_{p_0p_1c_1p_2x}&=\sum_{c_2}W_{p_0p_1c_1p_2c_2x}\,\phi_{\bq_2^b}(p_0,p_2c_2),\label{eq:K1}\\
A^{b}_{p_0p_1p_2x}&=e^{2\pi i\,\mathbf Q^{b}\cdot\br_{p_0}}\sum_{c_1}X^{b}_{p_0p_1c_1p_2x}\,\phi_{\bq_1^b}(p_0,p_1c_1),\label{eq:K2}
\end{align}
with $\mathbf Q^b=\bq_0^b+\bq_1^b+\bq_2^b$. $W$ is real, so Eq.~\eqref{eq:K1} is a real product with the stacked
real and imaginary parts of $\phi$. The intermediate $X$ has $B\,n_p^2n_s\,27$ elements and is processed in blocks
of $p_0$.

For the term in which both phase factors depend on the triplet, the two sums are merged. The products
$\phi_{\bq_1^b}(p_0,p_1c_1)\,\phi_{\bq_2^b}(p_0,p_2c_2)$ are formed first, and the sum over $(c_1,c_2)$ is then one
real matrix product of $W$ with their stacked real and imaginary parts for each $p_0$, without the intermediate
$X$. When the supercell is the primitive cell, the sum has a single term and reduces to an element-wise product.

phono3py averages the Fourier transform over the three choices of origin atom. When all shortest vectors in a
triple are unique ($\sigma_{p_0s_1s_2}=1$), these choices are equivalent and only the first is retained. We encode
this rule in $W^{(1)}=W(1+2\sigma)$ and $W^{(23)}=W(1-\sigma)$, apply Eqs.~\eqref{eq:K1}
and~\eqref{eq:K2} with cyclic permutations of the three legs, and combine
\begin{equation}
\tilde\Phi_{i l,j m,k n}=\tfrac13\bigl[A^{(1)}_{ijk,lmn}+A^{(2)}_{jik,mln}+A^{(3)}_{kji,nml}\bigr],
\end{equation}
where $i,j,k$ are atoms and $l,m,n$ Cartesian components. In the second and third terms one of the two phase
factors belongs to $\bq_0$, which is the same for all triplets of a grid point. The contraction with this factor is
done once per grid point, and each triplet requires only the contraction with the remaining factor.

\subsection{Projection onto normal modes}
\label{sec:proj}

The projected tensor
\begin{equation}
V_{\lambda_0\lambda_1\lambda_2}=\sum_{p_0p_1p_2}\sum_{\alpha\beta\gamma}
\frac{e_{p_0\alpha}(\lambda_0)e_{p_1\beta}(\lambda_1)e_{p_2\gamma}(\lambda_2)}{\sqrt{m_{p_0}m_{p_1}m_{p_2}}}
\tilde\Phi_{p_0\alpha,p_1\beta,p_2\gamma}
\label{eq:proj}
\end{equation}
is computed for all $n_b^3$ band triples of all $B$ triplets as three successive batched matrix products with
$E_t=\mathrm{diag}(m^{-1/2})\,[\mathbf e(\bq_t)]$, at a cost of $3n_b^4$ multiply--adds per triplet. The quantity
that is stored is $|V|^2/(\omega_0\omega_1\omega_2)$, and the remaining constants of Eq.~\eqref{eq:phinorm} and
$1/N_q$ are applied to the final sum. It is set to zero where any of the three frequencies is not above the
cut-off, with a mask.

\subsection{Branch-free tetrahedron integration with pruning}
\label{sec:thm}

For each band pair $(j_1,j_2)$ of a triplet and each of the 24 tetrahedra around $\bq_1$, the four vertex values of
$\omega_1+\omega_2$, $\omega_2-\omega_1$ and $\omega_1-\omega_2$ are sorted, $v_0\le v_1\le v_2\le v_3$, and the
weight of the central vertex is
\begin{equation}
w(\omega)=\frac16\sum_{\text{tet}}\sum_{r=1}^{3}\mathbf 1\bigl[v_{r-1}\le\omega<v_r\bigr]\,I_r(\omega)\,g_r(\omega),
\label{eq:thm}
\end{equation}
where $g_r$ and $I_r$ are the density of states and the vertex weight of region $r$ in the convention of
phonopy~\cite{Togo2023} and the factor $1/6$ is its normalisation of the 24 tetrahedra. We make five changes to
the evaluation. (a) The sort is the compare-and-swap network of phonopy written with element-wise minima and
maxima, and the position of the central vertex after sorting is the number of vertices below it. Among exactly
equal vertex values this position can differ from that of phonopy, which changes the weight only by rounding
because equal vertices have equal weights.
(b) The three region expressions are evaluated for whole arrays and selected with the masks of
Eq.~\eqref{eq:thm}, and every division is guarded by $|v_n-v_m|<10^{-10}$, as in phonopy. (c) A tetrahedron can
contribute only if $v_0\le\omega<v_3$. The set of (frequency, band pair, tetrahedron) elements that fulfil this
condition is formed first, $w$ is evaluated on that set only, and the results are added to the dense array.
(d) Before the vertex values are built, lower and upper bounds of the three combinations are obtained for each band
pair from the extreme vertex frequencies of the two bands, and the pairs whose bounds exclude all frequencies of
$\bq_0$ are discarded. (e) For cells with at least 30 branches the region expressions are compiled by the TorchScript
fuser of PyTorch into a small number of element-wise kernels.

\subsection{Occupation factors outside the band sums}
\label{sec:occ}

Let $g^{(1)}$ and $g^{(2)}$ be the weights of the first and of the bracketed $\delta$ functions of
Eq.~\eqref{eq:gamma} and $P^{(r)}_{j_0j_1j_2}=g^{(r)}|V|^2/(\omega_0\omega_1\omega_2)$. Then
\begin{multline}
\sum_{j_1j_2}\Bigl[(n_{j_1}+n_{j_2}+1)P^{(1)}+(n_{j_1}-n_{j_2})P^{(2)}\Bigr]\\
=\sum_{j_1}(n_{j_1}+1)S^{(1)}_{j_0j_1}+\sum_{j_2}n_{j_2}\bar S^{(1)}_{j_0j_2}
+\sum_{j_1}n_{j_1}S^{(2)}_{j_0j_1}-\sum_{j_2}n_{j_2}\bar S^{(2)}_{j_0j_2},
\label{eq:occ}
\end{multline}
with $S^{(r)}_{j_0j_1}=\sum_{j_2}P^{(r)}_{j_0j_1j_2}$ and $\bar S^{(r)}_{j_0j_2}=\sum_{j_1}P^{(r)}_{j_0j_1j_2}$,
which do not depend on temperature. The temperature-independent sums are formed once per triplet, reducing the cost for $N_T$ temperatures from
$\mathcal O(N_T n_b^3)$ to $\mathcal O(n_b^3+N_T n_b^2)$. Triplet multiplicities are retained. The identity
holds because $P$ vanishes for frequencies at or below the cutoff, for which phono3py skips the terms.

\subsection{Isotope scattering}
\label{sec:iso}

The isotope linewidth of mode $(\bq_0,j_0)$ follows Tamura~\cite{Tamura1983} and is evaluated as
\begin{equation}
\Gamma^{\mathrm{iso}}_{\bq_0j_0}=\frac{\pi\omega_{\bq_0j_0}^2}{4N_q}\sum_{g=1}^{N_q}\sum_{j}
w_{g,j_0j}\sum_{p} g_p\bigl|O_{g,j_0,j,p}\bigr|^2,\qquad
O_{g,j_0,j,p}=\sum_\alpha e^*_{p\alpha}(\bq_0j_0)\,e_{p\alpha}(\bq_gj),
\label{eq:iso}
\end{equation}
where $g$ runs over the $N_q$ points of the regular grid, $w_{g,j_0j}$ is the tetrahedron weight of
$\delta(\omega_{\bq_0j_0}-\omega_{\bq_gj})$, and $g_p=\sum_if_{ip}(1-m_{ip}/\bar m_p)^2$ is the mass variance of
atom $p$ with isotope fractions $f_{ip}$ and masses $m_{ip}$. The weights come from one call of the routine of
Sec.~\ref{sec:thm}, with the $n_b$ frequencies of $\bq_0$ as sampling points and all bands of all grid points as
vertices. The overlaps $O$ are formed for a block of grid points in one contraction, and the sum over atoms is a
product of $|O|^2$ with the vector $g_p$. The overlaps are needed only where $w_{g,j_0j}$ is non-zero, and when this
holds for fewer than 20\% of the band pairs they are formed for these elements only. Modes below the frequency
cut-off are removed with masks.

\subsection{Rules at vertices and for degenerate modes}
\label{sec:degen}

Two rules of recent phono3py versions are implemented as array operations. Since phono3py~4.7 a frequency that
equals a vertex of a tetrahedron receives the weight of the frequencies just above it instead of zero. This is the
half-open region of Eq.~\eqref{eq:thm}, and it matters mainly for the isotope term, where $\omega$ always equals the
central vertex of its own grid point. Since phono3py~4.8 the weights can be averaged over degenerate modes
with the option \texttt{--average-degenerate-weights}, which is off by default,
\begin{equation}
g^{(r)}_{j_0j_1j_2}\;\to\;\sum_{j_1'j_2'}(A_{\bq_1})_{j_1j_1'}\,g^{(r)}_{j_0j_1'j_2'}\,(A_{\bq_2})_{j_2j_2'},
\end{equation}
where $A_{\bq}$ averages over each set of modes within $10^{-4}$~THz at $\bq$. For the isotope term the weights are
averaged at $\bq_g$ and at the mode of interest, and $\Gamma^{\mathrm{iso}}$ is averaged over the degenerate set of the
latter. This reduces sensitivity of the linewidths to the eigenvector basis within degenerate subspaces
(Appendix~\ref{sec:gauge}). All results of this paper use both rules, and the reference calculations are phono3py~4.10
with the averaging option switched on. The behaviour of phono3py~3.12.1 (open regions, no averaging) can be
selected.

\subsection{Assembly of the Wigner conductivity}
\label{sec:wig}

Eq.~\eqref{eq:wigner} is evaluated without loops over temperatures, $\bq$ points or band pairs. The factor that
multiplies the velocities has shape $(N_T,N_{\mathrm{ir}},n_b,n_b)$ for $N_{\mathrm{ir}}$ irreducible points, the
velocity products summed over the star of $\bq$ have shape $(N_{\mathrm{ir}},n_b,n_b,6)$, and their product is
summed to give $\kappa$. The three-phonon linewidths are first averaged over the modes within $10^{-4}$~THz of
each other. A mask $|\omega_{\bq j}-\omega_{\bq j'}|<10^{-4}$~THz then splits the band pairs. The pairs inside the
mask, which are the diagonal pairs and the pairs of different but degenerate bands, form the populations
conductivity $\kP$, and all other pairs form the coherences conductivity $\kC$. For non-degenerate bands $\kP$
reduces to $\sum Cv^2\tau/(VN_q)$. Pairs that contain a mode below the frequency cut-off are set to zero.

\subsection{Memory and precision}
\label{sec:memory}

All steps operate on a batch of $B$ triplets of the same $\bq_0$. $B$ follows from a memory budget (by default
60\% of the free device memory) and the footprint per triplet,
\begin{equation}
b_{\mathrm{pp}}\approx (8b_{\mathrm{cx}}+80)n_b^3,\qquad
b_{\mathrm{thm}}\approx 6912\,n_b^2+96\,n_b^3\ \text{bytes},
\end{equation}
where $b_{\mathrm{cx}}=16$ bytes for complex128 and 8 bytes for complex64. The intermediate $X$ of Eq.~\eqref{eq:K1} and the set of active
elements of Sec.~\ref{sec:thm} have separate budgets. In the mixed-precision mode only the contractions of
Eqs.~\eqref{eq:K1}, \eqref{eq:K2} and \eqref{eq:proj} run in single precision, with single-precision copies of
the force constants, phase factors and eigenvectors. $|V|^2$ is formed in double precision, and the frequencies,
the tetrahedron weights, the sums of Eq.~\eqref{eq:occ}, the isotope term and the conductivity are always in
double precision. In the reported kernel tests, mixed precision changes the linewidths by approximately
$10^{-7}$ relative to double precision. For the ten cells of Table~\ref{tab:cpu} the conductivity changes by
$1.5\times10^{-8}$ to $1.2\times10^{-6}$.

\subsection{Host steps and force constants}
\label{sec:fc}

Reading the input files, choosing the mesh, and building the irreducible points, the triplets, the tetrahedron
neighbour tables and the shortest-vector tables are done once on the host with the phono3py Python API. No
interaction, integration-weight or conductivity routine of phono3py is called. The reordering of the third-order
force constants and their multiplication with the factors of the origin average are done on the GPU. When force constants are not supplied, they are fitted on the GPU from displacement and force data using
phono3py-compatible operations. Batched pseudoinverses replace many small independent solves, and only compact
rows of $\Phi^{(3)}$ are stored, reducing memory scaling from $n_s^3$ to $n_pn_s^2$.

\section{Validation details}\label{app:valid}
\label{sec:valid}

\subsection{Reference and protocol}

The reference is phono3py~4.10.0~\cite{Togo2023} with its default Rust backend (phonors~0.8.0). The Wigner
conductivity comes from the phono3py-wte plugin~0.1.0~\cite{wte}, in which one import statement was adapted to
phono3py~4.x. In this version a sampling frequency equal to a tetrahedron vertex receives the limit from above by
default. The averaging of the integration weights over degenerate modes
(\texttt{average\_degenerate\_weights}) is off by default and was switched on. All other options keep their
defaults, so \texttt{exclude\_gamma\_acoustic} and \texttt{symmetrize\_tetrahedra} are off. Our code is run with the
equivalent rules (Sec.~\ref{sec:degen}). Both codes read the same \texttt{fc2.hdf5}, \texttt{fc3.hdf5} and mesh, and
use the tetrahedron method with the RTA and the Wigner conductivity at $T=300$, 500 and 800~K.

Three sets of cells are compared (Table~\ref{tab:kappa}, with the meshes). The first set is the five stable
cells of Fig.~\ref{fig:valid}a: MgO, CuRhO$_2$ and AlCuO$_2$ with force constants from DFT, and Sc$_{11}$O$_9$ and
MgRh$_2$O$_4$ with force constants from a universal machine-learned potential. The second is TlRhO$_2$ (DFT), which
is dynamically unstable. It is also shown in Fig.~\ref{fig:valid}a and serves as a numerical test only. For
these six cells the second- and third-order force constants both come from a $2\times2\times2$ supercell. The third
set is the Si cells of Fig.~\ref{fig:valid}b, with force constants from the potential of the main text and no
pair cut-off for the third order. For the primitive cell the third-order force constants come from a 64-atom
supercell and the second-order ones from a 512-atom supercell. For the 64-site cells, with and without one Ge atom
or one vacancy, the third-order force constants come from the cell itself and the second-order ones from its
$2\times2\times2$ supercell. No non-analytic correction is used and no convergence study was made, so the absolute
conductivities are not converged predictions. This does not affect the comparison, because both codes receive
identical input. For the total conductivity, deviations are the maximum pointwise relative differences over the diagonal tensor
components and temperatures. For $\kP$, $\kC$ and phonon linewidths, deviations are the largest absolute
differences divided by the largest reference value of the corresponding quantity.

\subsection{Kernel tests}

Table~\ref{tab:unit} compares single steps with the corresponding routines of phono3py~4.10 on the CPU, at
selected $\bq$ points of MgO and CuRhO$_2$. The tests of the interaction, the linewidth and the isotope term use
the eigenvectors of phono3py as input, so that they do not depend on the choice of eigenvectors inside degenerate
sets. The three-phonon interaction and linewidth kernels agree to $2\times10^{-15}$ or better in double
precision. The velocity operators, which involve a finite difference, agree to $2\times10^{-10}$. The
mixed-precision mode is a separate test and changes the linewidths by $10^{-7}$.

\begin{table}[b]
\caption{Kernel tests against phono3py~4.10 (MgO and CuRhO$_2$ unless noted). Deviations are relative as defined
in the text, except for the frequencies, where the largest absolute difference is given. ``Same $\mathbf e$''
means that the eigenvectors of phono3py were supplied.}
\label{tab:unit}
\begin{ruledtabular}
\begin{tabular}{lc}
quantity & deviation\\
\hline
$D(\bq)$ vs phonopy & $\le10^{-12}$\\
$\omega_{\bq j}$ ($|\omega|>10^{-3}$~THz) & $\le1.5\times10^{-13}$~THz\\
$|\Phi|^2$, all triplets at one $\bq$ & $\le2\times10^{-15}$\\
$\Gamma$ (tetrahedron, same $\mathbf e$) & $\le2\times10^{-15}$\\
$\Gamma^{\mathrm{iso}}$ (same $\mathbf e$) & $\le3\times10^{-16}$\\
$\mathbf v$, $\sum_{\mathcal S}\mathbf v\otimes\mathbf v^*$ & $\le2\times10^{-10}$\\
$\Gamma$, Gaussian smearing ($\sigma=0.2$~THz) & $1.7\times10^{-15}$\\
$\kP$, $\kC$, Gaussian smearing\footnote{$\sigma=0.1$~THz, full calculation for MgO on a $20^3$ mesh.} & $5.7\times10^{-13}$\\
$\Gamma$, mixed vs.\ double precision\footnote{Our code against itself.} & $10^{-7}$\\
$\Phi^{(2)},\Phi^{(3)}$ from forces\footnote{13 other cells with 32--512 atoms, against the force constants of phono3py.} & $\le2.3\times10^{-14}$\\
\end{tabular}
\end{ruledtabular}
\end{table}

\subsection{Full conductivity}

In the full calculation our code uses its own eigenvectors. Table~\ref{tab:kappa} lists the results on the GPU.
\begin{itemize}
\item \emph{Without isotope scattering.} For all stable cells the total conductivity $\kappa=\kP+\kC$ agrees to
  $5\times10^{-10}$ or better (largest $|\kappa/\kappa_{\mathrm{ref}}-1|$ over the diagonal components and the three
  temperatures), and $\kP$ and $\kC$ separately to $3\times10^{-10}$.
  The linewidths agree to $3\times10^{-10}$, except for a few nearly degenerate modes of AlCuO$_2$
  ($3\times10^{-8}$).
\item \emph{With isotope scattering.} The total conductivity agrees to $3.4\times10^{-5}$ or better for the five
  stable cells of Fig.~\ref{fig:valid}a and to $5.2\times10^{-5}$ for the Si cells. The separate deviations of $\kP$
  and $\kC$ reach $3.4\times10^{-5}$ and $6.0\times10^{-5}$, while the three-phonon linewidths are
  unchanged. The residual comes from $\Gamma^{\mathrm{iso}}$ (Sec.~\ref{sec:isoresid}).
\item \emph{Unstable TlRhO$_2$.} Our code stops when the lowest frequency on the mesh is below $-0.1$~THz. For
  this test the check was switched off. The modes below the frequency cut-off are then excluded as in phono3py.
  Both codes give the same lowest frequency, $-0.9267$~THz, the total conductivity agrees to $5\times10^{-7}$, and single linewidths differ by up to $6\times10^{-4}$.
\end{itemize}

% Full-conductivity validation table (reverified 2026-11-24, RTX PRO 6000).
\begin{table*}[t]
\caption{This work on the GPU (RTX PRO 6000) vs.\ phono3py~4.10 + phono3py-wte (same inputs and
mesh, tetrahedron method, RTA+Wigner, degenerate averaging). $\kP^{xx}$ in W\,m$^{-1}$K$^{-1}$ at 300~K, rounded to
five decimals. All deviations are computed from the unrounded values over $xx,yy,zz$ and $T=300$, 500, 800~K. The
deviation of the total conductivity $\kappa=\kP+\kC$ is the largest $|\kappa/\kappa_{\mathrm{ref}}-1|$. The deviations
of $\kP$ and $\kC$ are the largest absolute difference divided by the largest reference value. $\Gamma$ is the
three-phonon linewidth without the isotope term, compared over all modes and $T$ relative to $\max\Gamma$.}
\label{tab:kappa}
\begin{ruledtabular}
\begin{tabular}{llccccccc}
case & isotopes & mesh & $\kP^{xx}$ phono3py & $\kP^{xx}$ this work & dev.\ $\kappa$ & dev.\ $\kP$ & dev.\ $\kC$ & dev.\ $\Gamma$\\
\hline
MgO & no & $20^3$ & 182.89844 & 182.89843 & $4.3\times10^{-10}$ & $3.0\times10^{-10}$ & $5.4\times10^{-11}$ & $2.7\times10^{-10}$\\
MgO & yes & $20^3$ & 138.15566 & 138.15098 & $3.4\times10^{-5}$ & $3.4\times10^{-5}$ & $1.3\times10^{-5}$ & $2.7\times10^{-10}$\\
CuRhO$_2$ & no & $19^3$ & 29.08484 & 29.08484 & $1.8\times10^{-10}$ & $1.9\times10^{-10}$ & $6.7\times10^{-12}$ & $3.0\times10^{-10}$\\
CuRhO$_2$ & yes & $19^3$ & 28.42091 & 28.42093 & $8.4\times10^{-7}$ & $5.9\times10^{-7}$ & $2.2\times10^{-6}$ & $3.0\times10^{-10}$\\
AlCuO$_2$ & no & $20^3$ & 39.25301 & 39.25301 & $1.9\times10^{-10}$ & $1.0\times10^{-10}$ & $9.1\times10^{-12}$ & $2.9\times10^{-8}$\\
AlCuO$_2$ & yes & $20^3$ & 38.38441 & 38.38442 & $4.3\times10^{-7}$ & $4.2\times10^{-7}$ & $1.1\times10^{-7}$ & $2.9\times10^{-8}$\\
Sc$_{11}$O$_9$ & no & $11{\times}7{\times}6$ & 5.61517 & 5.61517 & $2.6\times10^{-12}$ & $2.5\times10^{-12}$ & $1.2\times10^{-12}$ & $1.6\times10^{-12}$\\
Sc$_{11}$O$_9$ & yes & $11{\times}7{\times}6$ & 5.60949 & 5.60949 & $2.6\times10^{-12}$ & $2.5\times10^{-12}$ & $1.2\times10^{-12}$ & $1.6\times10^{-12}$\\
MgRh$_2$O$_4$ & no & $6^3$ & 14.87862 & 14.87862 & $1.3\times10^{-12}$ & $1.3\times10^{-12}$ & $6.0\times10^{-12}$ & $1.7\times10^{-12}$\\
MgRh$_2$O$_4$ & yes & $6^3$ & 14.80482 & 14.80475 & $4.5\times10^{-6}$ & $4.7\times10^{-6}$ & $4.7\times10^{-6}$ & $1.7\times10^{-12}$\\
\hline
\multicolumn{9}{l}{Dynamically unstable, stability check switched off}\\
TlRhO$_2$ & no & $18^3$ & 2.89753 & 2.89753 & $3.8\times10^{-7}$ & $2.1\times10^{-9}$ & $6.6\times10^{-7}$ & $6.3\times10^{-4}$\\
TlRhO$_2$ & yes & $18^3$ & 2.88190 & 2.88190 & $4.7\times10^{-7}$ & $4.8\times10^{-7}$ & $6.6\times10^{-7}$ & $6.3\times10^{-4}$\\
\hline
\multicolumn{9}{l}{Si cells}\\
Si & no & $19^3$ & 132.36501 & 132.36501 & $3.0\times10^{-12}$ & $2.7\times10^{-12}$ & $1.2\times10^{-11}$ & $8.0\times10^{-12}$\\
Si & yes & $19^3$ & 124.13909 & 124.13804 & $8.3\times10^{-6}$ & $8.4\times10^{-6}$ & $4.1\times10^{-5}$ & $8.0\times10^{-12}$\\
Si$_{64}$ & no & $6^3$ & 132.25325 & 132.25325 & $4.3\times10^{-11}$ & $3.3\times10^{-11}$ & $3.3\times10^{-11}$ & $9.7\times10^{-11}$\\
Si$_{64}$ & yes & $6^3$ & 124.03264 & 124.03112 & $2.1\times10^{-5}$ & $1.2\times10^{-5}$ & $6.0\times10^{-5}$ & $9.7\times10^{-11}$\\
Si$_{63}$Ge & no & $6^3$ & 56.59890 & 56.59890 & $4.9\times10^{-13}$ & $2.8\times10^{-13}$ & $1.1\times10^{-12}$ & $1.0\times10^{-12}$\\
Si$_{63}$Ge & yes & $6^3$ & 53.60484 & 53.60183 & $5.2\times10^{-5}$ & $5.6\times10^{-5}$ & $2.6\times10^{-5}$ & $1.0\times10^{-12}$\\
Si$_{63}$V$_{\mathrm{Si}}$ & no & $6^3$ & 6.30593 & 6.30593 & $5.3\times10^{-12}$ & $7.4\times10^{-12}$ & $8.7\times10^{-13}$ & $2.8\times10^{-12}$\\
Si$_{63}$V$_{\mathrm{Si}}$ & yes & $6^3$ & 6.09919 & 6.09922 & $7.5\times10^{-6}$ & $4.5\times10^{-6}$ & $2.0\times10^{-5}$ & $2.8\times10^{-12}$\\
\end{tabular}
\end{ruledtabular}
\end{table*}

\subsection{Degenerate modes and vertex coincidences}
\label{sec:gauge}

Within a degenerate phonon subspace, eigenvectors are not uniquely defined, and the distribution of
mode-resolved three-phonon interactions can depend on the chosen basis. Numerical differences can also arise when
a sampling frequency coincides with a tetrahedron vertex. To match phono3py~4.10, Phono3GPU uses the same vertex
convention and enables averaging of integration weights over degenerate modes (Sec.~\ref{sec:degen}). Full
conductivity agreement is reported in Table~\ref{tab:kappa}. The remaining numerical sensitivity of isotope
scattering is discussed in Sec.~\ref{sec:isoresid}.

\subsection{Numerical sensitivity of isotope scattering}
\label{sec:isoresid}

With both rules, $\Gamma^{\mathrm{iso}}$ still differs from the reference at some modes. For CuRhO$_2$ on a $10^3$
mesh (116 irreducible points, 1392 modes) 67 modes differ by more than $10^{-6}$ of the largest
$\Gamma^{\mathrm{iso}}$, and the largest difference is 8.7\% of that value. At these modes the sampling frequency
coincides, to about $10^{-15}$~THz, with two nearly degenerate vertices, so the tetrahedron region that is selected
depends on rounding. The effect on the conductivity is small, a relative change of $6.5\times10^{-6}$ in $\kP$. The same sensitivity appears
inside the reference code. The C and Rust backends of phono3py~4.10, compared with the same input and without
averaging, differ at 404 of the 1392 modes by the same criterion, with the same largest difference and a relative
change of $8.8\times10^{-5}$ in $\kP$. In the same runs the three-phonon linewidths agree to $3\times10^{-10}$, and for Sc$_{11}$O$_9$ the conductivity
with isotope scattering agrees to $3\times10^{-12}$ as well.

\section{Performance details}\label{app:perf}\label{sec:bench}

\subsection{Wall time and memory}

All benchmarks report total wall time from process start to exit, including input and output. The CPU reference was phono3py~4.10 (Rust backend) with phono3py-wte on a dedicated node with two Intel Xeon Gold 6230 processors (40 cores, 187~GB), using 32 OpenMP threads. Phono3GPU used one RTX A6000 (48~GB) or RTX PRO 6000 Blackwell (96~GB) with four host CPU cores. The inputs, meshes and isotope settings were matched.

Table~\ref{tab:cpu} lists the wall times and speedups plotted in Fig.~\ref{fig:valid}b. GPU execution was not always faster for small cells without isotope scattering, as shown by MgO (6.8~s on the CPU versus 8.3~s on Blackwell). Peak CPU memory was 0.07--0.14~GB for the small cells, 1.0~GB for Sc$_{11}$O$_9$, 11~GB for MgRh$_2$O$_4$ and 13--14~GB for 64-site Si. For MgRh$_2$O$_4$ and 64-site Si, peak device memory was 20 and 16~GB on the A6000, compared with 29 and 31~GB on Blackwell.

\begin{table*}[t]
\caption{Total wall times for RTA+Wigner conductivity with tetrahedron integration and natural isotope scattering. The CPU reference is phono3py~4.10 with phono3py-wte (32 threads). GPU runs use four host CPU cores and one GPU. Speedups ($s$) are CPU time divided by GPU time, reported consistently with Fig.~\ref{fig:valid}b. MgRh$_2$O$_4$ uses the 56-atom conventional cell. TlRhO$_2$ is dynamically unstable and was timed with the stability check disabled.}
\label{tab:cpu}
\begin{ruledtabular}
\begin{tabular}{lccrrrrr}
case & $N_{\mathrm{at}}$ & mesh & CPU (s) & A6000 (s) & $s_{\mathrm A}$ & Blackwell (s) & $s_{\mathrm B}$\\
\hline
MgO & 2 & $20^3$ & 38.4 & 10.9 & 3.5 & 8.6 & 4.5\\
CuRhO$_2$ & 4 & $19^3$ & 127.6 & 42.3 & 3.0 & 40.8 & 3.1\\
AlCuO$_2$ & 4 & $20^3$ & 171.6 & 52.0 & 3.3 & 55.7 & 3.1\\
TlRhO$_2$ & 4 & $18^3$ & 100.3 & 31.9 & 3.1 & 37.2 & 2.7\\
Sc$_{11}$O$_9$ & 20 & $11{\times}7{\times}6$ & 606.7 & 59.6 & 10.2 & 35.3 & 17.2\\
MgRh$_2$O$_4$ & 56 & $6^3$ & 563.9 & 27.7 & 20.4 & 13.7 & 41.2\\
Si (primitive) & 2 & $19^3$ & 40.4 & 12.7 & 3.2 & 8.8 & 4.6\\
Si$_{64}$ & 64 & $6^3$ & 574.5 & 32.7 & 17.6 & 17.0 & 33.8\\
Si$_{63}$Ge & 64 & $6^3$ & 564.2 & 28.4 & 19.9 & 15.2 & 37.1\\
Si$_{63}$V$_{\mathrm{Si}}$ & 63 & $6^3$ & 531.4 & 31.4 & 16.9 & 14.3 & 37.2\\
\end{tabular}
\end{ruledtabular}
\end{table*}

\subsection{Comparison with the GPU plug-in for phono3py}
\label{sec:p3gpu}

The plug-in~\cite{phono3pygpu} (commit 4bf2777 of 24 November 2026, with CuPy) was run on the same nodes with the same inputs, meshes and isotope settings, one GPU and four host CPU cores. It was run with its default settings and with larger batch settings, which changed its times by 5 to 10\%, and the shorter time is listed in Table~\ref{tab:p3gpu}. Its conductivities agree with the CPU reference within $2\times10^{-5}$.

\begin{table*}[t]
\caption{Total wall times (s) of the GPU plug-in for phono3py and of Phono3GPU for the calculations of Table~\ref{tab:cpu}. The ratio is the time of the plug-in divided by the time of Phono3GPU, as plotted in Fig.~\ref{fig:valid}c for Blackwell.}
\label{tab:p3gpu}
\begin{ruledtabular}
\begin{tabular}{lrrrrrr}
 & \multicolumn{3}{c}{RTX A6000} & \multicolumn{3}{c}{RTX PRO 6000 Blackwell}\\
case & plug-in & this work & ratio & plug-in & this work & ratio\\
\hline
MgO & 86.3 & 10.9 & 7.9 & 56.1 & 8.6 & 6.5\\
CuRhO$_2$ & 293.1 & 42.3 & 6.9 & 212.1 & 40.8 & 5.2\\
AlCuO$_2$ & 391.1 & 52.0 & 7.5 & 286.7 & 55.7 & 5.1\\
TlRhO$_2$ & 217.5 & 31.9 & 6.8 & 156.5 & 37.2 & 4.2\\
Sc$_{11}$O$_9$ & 198.5 & 59.6 & 3.3 & 85.7 & 35.3 & 2.4\\
MgRh$_2$O$_4$ & 103.5 & 27.7 & 3.7 & 45.5 & 13.7 & 3.3\\
Si (primitive) & 66.8 & 12.7 & 5.3 & 44.1 & 8.8 & 5.0\\
Si$_{64}$ & 122.1 & 32.7 & 3.7 & 59.9 & 17.0 & 3.5\\
Si$_{63}$Ge & 74.9 & 28.4 & 2.6 & 34.8 & 15.2 & 2.3\\
Si$_{63}$V$_{\mathrm{Si}}$ & 69.3 & 31.4 & 2.2 & 28.6 & 14.3 & 2.0\\
\end{tabular}
\end{ruledtabular}
\end{table*}

\subsection{Time of the individual steps}

Table~\ref{tab:stages} splits one run on the A6000 into its steps. For the two- and four-atom cells the tetrahedron weights take as long as the three-phonon interaction or longer, and for CuRhO$_2$ the two steps together take 85\% of the run. For the 56-atom cell the interaction is five times more expensive than the weights. This is the reason why the gain over the CPU grows with the number of atoms: the cost of the interaction grows faster with the number of bands than that of the weights, and it consists of dense contractions.

\begin{table}[t]
\caption{Time of the steps of one run on the RTX A6000 in seconds (natural isotopes, double precision). The device was synchronised after each step, so the totals are larger than the wall times of Table~\ref{tab:cpu}. The remaining linewidth steps are the triplet search on the CPU, the occupation factors and the band sums.}
\label{tab:stages}
\begin{ruledtabular}
\begin{tabular}{lccc}
step & MgO & CuRhO$_2$ & MgRh$_2$O$_4$\\
\hline
Phonons and velocity operators & 0.2 & 0.7 & 4.1\\
Three-phonon interaction & 2.1 & 34.8 & 71.1\\
Tetrahedron weights & 2.9 & 40.1 & 13.5\\
Remaining linewidth steps & 1.3 & 6.3 & 12.4\\
Isotope linewidth & 1.0 & 4.2 & 3.3\\
Input, setup, conductivity and output & 1.5 & 1.7 & 65.5\\
\hline
Total & 9.0 & 87.8 & 169.9\\
\end{tabular}
\end{ruledtabular}
\end{table}

\subsection{Mixed precision and larger defect cells}

On Blackwell, mixed precision reduced the compute time (phonon solution and conductivity, excluding input, force-constant preparation and output) by factors of 2.7 for MgRh$_2$O$_4$ and 1.8 for Sc$_{11}$O$_9$, while the conductivity changed by $1\times10^{-7}$ and $5\times10^{-8}$ relative to double precision (Table~\ref{tab:mixed}).

\begin{table}[t]
\caption{Mixed precision against double precision on the RTX PRO 6000 Blackwell (natural isotopes). The compute time covers the phonon solution and the conductivity, and the total wall time the whole run. The deviations of the mixed-precision results from the double-precision ones are defined as in Table~\ref{tab:kappa}.}
\label{tab:mixed}
\begin{ruledtabular}
\begin{tabular}{lcc}
 & Sc$_{11}$O$_9$ & MgRh$_2$O$_4$\\
\hline
Compute time, double (s) & 49.0 & 28.3\\
Compute time, mixed (s) & 27.5 & 10.6\\
Total wall time, double (s) & 56.5 & 31.4\\
Total wall time, mixed (s) & 35.3 & 13.7\\
Deviation of $\kappa$ & $5.4\times10^{-8}$ & $9.8\times10^{-8}$\\
Deviation of $\kP$ & $5.8\times10^{-8}$ & $1.0\times10^{-7}$\\
Deviation of $\kC$ & $1.1\times10^{-8}$ & $7.9\times10^{-9}$\\
Deviation of $\Gamma$ & $9.5\times10^{-8}$ & $3.7\times10^{-7}$\\
\end{tabular}
\end{ruledtabular}
\end{table}

For 128-site Si supercells on the A6000, two Ge substitutions (128 atoms, 384 branches, 34 irreducible $\bq$ points) took 40~min, while two vacancies (126 atoms, 378 branches, 56 irreducible points) took 74~min, both on a $6\times6\times3$ mesh with approximately 21~GB peak device memory. These timings demonstrate feasibility and are not CPU speedup measurements.

% =====================================================================
\section{Atom-resolved linewidths}\label{si:atom}

The three-phonon vertex is partitioned by the first atomic index of the third-order force constants, which is
the index contracted with the eigenvector of the mode $\lambda$ itself,
$\Phi_{\lambda\lambda'\lambda''}=\sum_i\Phi^{(i)}_{\lambda\lambda'\lambda''}$.
Replacing $|\Phi|^2$ in Eq.~\eqref{eq:gamma} with
$\operatorname{Re}[\Phi^{(i)*}\Phi]$, using the same tetrahedron weights, occupations and degenerate-mode
averaging, gives the partial linewidth $\Gamma^{(3)}_{\lambda,i}$, where $\Gamma^{(3)}_\lambda$ denotes the
three-phonon linewidth of Eq.~\eqref{eq:gamma}.
The isotope term is assigned directly to the atomic contributions in Eq.~\eqref{eq:iso}.
Thus, with $\Gamma_{\lambda,i}=\Gamma^{(3)}_{\lambda,i}+\Gamma^{\mathrm{iso}}_{\lambda,i}$,
$\sum_i\Gamma_{\lambda,i}=\Gamma^{(3)}_\lambda+\Gamma^{\mathrm{iso}}_\lambda$.
The relative sum-rule residual is at most $1.4\times10^{-15}$ in the reported calculations.
Because the decomposition retains interference between atomic contributions, individual
$\Gamma_{\lambda,i}$ may be negative and should not be interpreted as independent local scattering rates.

We characterise the atom-resolved contribution of heat-carrying modes and the first-order
sensitivity of the RTA populations conductivity by
\begin{equation}
\Gamma_i=\frac{\sum_\lambda C_\lambda v_\lambda^2\,\Gamma_{\lambda,i}}{\sum_\lambda C_\lambda v_\lambda^2},\qquad
S_i^{\mathrm P}=\sum_\lambda\kappa_{\lambda}^{\mathrm P}\frac{\Gamma_{\lambda,i}}{\Gamma_\lambda},
\qquad \sum_i S_i^{\mathrm P}=\kP,
\end{equation}
where $v_\lambda^2=|\mathbf v_\lambda|^2/3$ and $\kappa_{\lambda}^{\mathrm P}$ denotes the RTA
mode contribution. For the calculations here, $\Gamma_\lambda$ contains three-phonon and isotope
scattering, without a separate boundary-scattering term. Formally scaling the assigned
$\Gamma_{\lambda,i}$ by $1-\epsilon$ at fixed harmonic properties changes
$\kP$ by $\epsilon S_i^{\mathrm P}+\mathcal O(\epsilon^2)$.
This is not a prediction for the response to a physical perturbation of atom $i$,
which would generally change other terms as well.
The Wigner coherence contribution $\kC$ is not included in $S_i^{\mathrm P}$.
In pristine Si, symmetry-equivalent atoms have equal $\Gamma_i$, providing
a reference for the spatial redistribution in defective cells.

The analysis additionally outputs atom-projected mode weights from the phonon eigenvectors.
It is run separately from the production conductivity calculation and requires
150--165~s and approximately 19~GB of GPU memory for the 64-site Si cells on the RTX A6000,
compared with approximately two minutes for the conductivity calculation.
Previous atom-resolved conductivity analyses have used harmonic or molecular-dynamics
formulations~\cite{Gautam2024,Yokoi2025}. Here the decomposition resolves the
anharmonic BTE linewidth into atomic contributions while retaining their interference.

\section{Direct solution of the linearised BTE}\label{app:lbte}

In addition to the relaxation-time approximation (RTA), Phono3GPU solves the linearised phonon Boltzmann transport equation using the symmetrised collision operator of phono3py~\cite{Chaput2013}. The populations conductivity is
\begin{equation}
\begin{aligned}
\kP^{\alpha\beta}&=\frac{2k_BT^2}{VN_q}
\sum_{\lambda\lambda'}X^\alpha_\lambda
(\Omega^+)_{\lambda\lambda'}X^\beta_{\lambda'},\\
X^\alpha_\lambda&=\frac{\hbar\omega_\lambda v^\alpha_\lambda}
{4k_BT^2\sinh(\hbar\omega_\lambda/2k_BT)},
\end{aligned}
\label{eq:lbte}
\end{equation}
with the collision operator
\begin{multline}
\Omega_{\lambda\lambda'}=\delta_{\lambda\lambda'}\Gamma_\lambda
+\frac{18\pi}{\hbar^2}\sum_{\lambda''}
\frac{|\Phi_{\lambda\lambda'\lambda''}|^2}
{\sinh(\hbar\omega_{\lambda''}/2k_BT)}\\
\times\bigl[\delta(\omega_\lambda-\omega_{\lambda'}-\omega_{\lambda''})
+\delta(\omega_\lambda+\omega_{\lambda'}-\omega_{\lambda''})\\
+\delta(\omega_\lambda-\omega_{\lambda'}+\omega_{\lambda''})\bigr].
\label{eq:colmat}
\end{multline}
Here $\Phi$ follows the convention of Eq.~\eqref{eq:phinorm}, and $\Omega^+$ denotes the pseudo-inverse obtained by discarding eigenvalues below $10^{-8}$, as in phono3py. When the populations conductivity is obtained from the direct LBTE solution, only the coherence contribution defined in Sec.~\ref{sec:wig} is retained from Eq.~\eqref{eq:wigner}.

\subsection{GPU assembly and direct solution}\label{sec:lbte}

The irreducible collision matrix has indices $(i_0j_0\alpha,ij\beta)$ and dimension $N=3N_{\mathrm{ir}}n_b$. Its rows are constructed in the same triplet batches used for the three-phonon linewidths. The tetrahedron channel for Eq.~\eqref{eq:colmat} reuses the integration routine of Sec.~\ref{sec:thm}. With $P^{(3)}=g^{(3)}|\Phi|^2$ and $s_j(\bq,T)=1/\sinh(\hbar\omega_{\bq j}/2k_BT)$, the temperature dependence is applied to two partial sums,
\begin{equation}
\begin{aligned}
A_{T,b,j_0j_1}&=\sum_{j_2}P^{(3)}_{b,j_0j_1j_2}s_{j_2}(\bq_2,T),\\
B_{T,b,j_0j_2}&=\sum_{j_1}P^{(3)}_{b,j_0j_1j_2}s_{j_1}(\bq_1,T).
\end{aligned}
\label{eq:lbte_partial}
\end{equation}
The two sums account for both assignments of the remaining triplet points to collision-matrix columns. The resulting blocks $M_{\bq',j_0j}$ are gathered using phono3py's triplet map and combined over the point-group rotations $R$ in one batched contraction, which gives the off-diagonal part of the collision matrix,
\begin{equation}
\Omega^{\mathrm{off}}_{(i_0j_0\alpha),(ij\beta)}=
\sum_R M_{R\bq_i,j_0j}R_{\alpha\beta}.
\label{eq:lbte_assembly}
\end{equation}
Diagonal linewidth terms, star weights and degenerate-mode averaging are then applied as block operations, followed by symmetrisation of $\Omega$. For each temperature the symmetric eigensolver computes $\Omega=U\operatorname{diag}(w)U^{\mathsf T}$ and $\Omega^+X=U\operatorname{diag}(w^+)U^{\mathsf T}X$ on the GPU. Collision matrices are staged in pinned host memory and processed one temperature at a time when necessary to fit the device memory budget. The dense eigendecomposition scales as $\mathcal O(N^3)$.

Before the decomposition the matrix is reduced to the subspace allowed by symmetry. For an irreducible point $\bq_i$, Eq.~\eqref{eq:lbte_assembly} contains only Cartesian vectors that are invariant under the $n_i$ rotations that leave $\bq_i$ unchanged. The matrix is therefore rotated to the eigenvectors of the projectors $P_i=n_i^{-1}\sum_{R\bq_i=\bq_i}R$, and only the block of the invariant vectors is decomposed. The removed directions are null vectors of $\Omega$ and do not contribute to $\Omega^+$, so the solution changes only by rounding (at most $2\times10^{-14}$ in $\kP^{\mathrm{exact}}$ for the cells tested). The reduction of $N$ is a factor of three for the 64-site Si cells and MgRh$_2$O$_4$, 26\% for MgO and 10\% for CuRhO$_2$ on the meshes of Table~\ref{tab:cpu}. The step is skipped when the reduction is below 5\%, as for Sc$_{11}$O$_9$.

\subsection{Numerical validation}\label{sec:val_lbte}

Table~\ref{tab:lbte} compares the direct LBTE populations conductivity, Wigner coherence contribution and linewidths with phono3py~4.10 and phono3py-wte. Without isotope scattering the largest normalised difference in $\kP^{\mathrm{exact}}$ is $3.9\times10^{-10}$. With isotope scattering the difference on the full meshes is at most $3.3\times10^{-5}$ (MgO, $20^3$), which reflects the numerical sensitivity of the isotope tetrahedron weights near coincident vertices (Sec.~\ref{sec:isoresid}). In the examples shown, the direct solution increases $\kP^{xx}$ relative to the RTA by approximately 20\% for CuRhO$_2$ on a $10^3$ mesh and 1.6\% for MgO on an $8^3$ mesh.

\begin{table*}[t]
\caption{Direct LBTE populations conductivity and Wigner coherence compared with phono3py~4.10 and phono3py-wte. Values of $\kP^{\mathrm{exact}}$ and the RTA contribution are for $xx$ at 300~K in W\,m$^{-1}$K$^{-1}$. Deviations are maxima over $xx$, $yy$, $zz$ and 300, 500 and 800~K. Isotope scattering is excluded except for Sc$_{11}$O$_9$, for which no isotope-free LBTE reference was computed.}
\label{tab:lbte}
\begin{ruledtabular}
\begin{tabular}{llcccccc}
case & mesh & $N$ & $\kappa_{\mathrm{P},xx}^{\mathrm{exact}}$ (reference) & dev. $\kP^{\mathrm{exact}}$ & dev. $\kC$ & dev. $\Gamma$ & $\kappa_{\mathrm{P},xx}^{\mathrm{RTA}}$\\
\hline
MgO & $8^3$ & 522 & 246.3117 & $3.3\times10^{-10}$ & $1.1\times10^{-10}$ & $4.3\times10^{-10}$ & 242.54\\
CuRhO$_2$ & $10^3$ & 4176 & 32.7490 & $6.8\times10^{-11}$ & $5.0\times10^{-11}$ & $2.9\times10^{-10}$ & 27.34\\
MgO & $20^3$ & 4608 & 187.5665 & $2.4\times10^{-10}$ & $5.4\times10^{-11}$ & $2.7\times10^{-10}$ & 182.90\\
CuRhO$_2$ & $19^3$ & 24\,120 & 34.6400 & $7.7\times10^{-11}$ & $6.7\times10^{-12}$ & $3.0\times10^{-10}$ & 29.08\\
AlCuO$_2$ & $20^3$ & 28\,116 & 41.368 & $3.9\times10^{-10}$ & $9.1\times10^{-12}$ & $2.9\times10^{-8}$ & 39.25\\
Sc$_{11}$O$_9$ & $11{\times}7{\times}6$ & 41\,760 & 5.697 & $2.2\times10^{-12}$ & $1.2\times10^{-12}$ & $1.6\times10^{-12}$ & 5.61\\
\end{tabular}
\end{ruledtabular}
\end{table*}

\subsection{Computational cost}\label{sec:lbte_cost}

The direct solution increases the cost of the CPU reference substantially. With natural isotopes, the conductivity calculation of phono3py~4.10 on 32 threads took 380~s for MgO ($20^3$), $3.58\times10^4$~s for CuRhO$_2$ ($19^3$), $6.03\times10^4$~s for AlCuO$_2$ ($20^3$) and $1.68\times10^5$~s for Sc$_{11}$O$_9$ ($11\times7\times6$). The corresponding compute times of Phono3GPU on the RTX PRO 6000 Blackwell were 8.9, 196, 294 and 762~s. Both sets of times cover the conductivity calculation only, without reading the input and writing the output, and they are for the direct LBTE with isotopes rather than the RTA benchmarks of Appendix~\ref{app:perf}. The speedup arises from reusing the three-phonon triplet batches to assemble $\Omega$ and solving the dense eigenproblem on the GPU. The largest case, Sc$_{11}$O$_9$ with $N=41\,760$, used approximately 85~GB of device memory, with temperature-dependent matrices staged through host memory.

\section{Comparison with other GPU codes}
\label{app:othercodes}

% Optional neutral disclosure (the GPU_PBTE numbers come from a copy with one added initialisation):
% GPU\_PBTE was run with one added initialisation of a device array.

FourPhonon\_GPU, GPU\_PBTE, Phoebe and $\kappa$ALDo were run on the same RTX A6000 nodes with four host cores. These
codes use Gaussian smearing, so the comparison is made with our Gaussian option ($\sigma=0.1$~THz) without isotope
scattering, and not with the tetrahedron method of the main text. The averaging over the choice of the origin atom,
which phono3py applies by default (Sec.~\ref{sec:fc3rec}), is not used by the other codes and changes $\kappa$ by up
to 53\% (MgO). Our code was therefore run without it for this comparison. Against this reference, $\kP^{xx}$ at 300~K
differs by $+0.2$\% (MgO) and $+0.5$\% (CuRhO$_2$) for GPU\_PBTE, by $-3.4$\% and $+0.1$\% for Phoebe, by $-2.0$\% (MgO)
for FourPhonon, which uses adaptive smearing, and by $+0.3$\% (MgO) for $\kappa$ALDo.

Table~\ref{tab:othercodes} lists the total wall times. Only the cases for which a code gave a usable conductivity are
included. For MgO our code and GPU\_PBTE take the same time. For the four-atom cells our code is 20 to 90 times
faster than Phoebe and GPU\_PBTE. With all 32 host cores instead of four, Phoebe is 1.2 to 1.5 times faster (MgO 75.8
to 61.9~s, CuRhO$_2$ 641 to 525~s), so the number of host cores does not explain the difference.

These times should be read with care. The other codes were built by us from their public sources with compilers and
libraries that differ from those of their authors, their inputs were converted by us from the phono3py force
constants, and they were run with default settings without tuning for each code. Part of the differences may
therefore come from our builds and settings rather than from the codes themselves.

\begin{table}[h]
\caption{Total wall time (s) of one run on an RTX~A6000 with four host cores (Gaussian smearing, no isotopes).
This work, Phoebe, GPU\_PBTE and $\kappa$ALDo compute 300, 500 and 800~K in one run, and FourPhonon is run once per
temperature (300~K shown). GPU\_PBTE runs in four separate stages, and their sum is given. Only cases with a usable
conductivity from the respective code are listed.}
\label{tab:othercodes}
\begin{ruledtabular}
\begin{tabular}{lccccc}
case & this work & Phoebe & GPU\_PBTE & FourPhonon & $\kappa$ALDo\\
\hline
MgO & 27.4 & 75.8 & 27.0 & 89.5 & 4581\\
CuRhO$_2$ & 27.7 & 641 & 1622 & -- & --\\
TlRhO$_2$ & 20.6 & 530 & 1887 & -- & --\\
\end{tabular}
\end{ruledtabular}
\end{table}

\section{Further analysis of the defects in Si}\label{si:pairs}

\begin{table*}[t]
\caption{
Defect cells of Fig.~\ref{fig:defect} at 300~K with natural isotopes.
The total thermal conductivity $\kappa=\kP+\kC$ is averaged over
the three Cartesian directions.
The factor denoted by $\sum Cv^2$ is defined as
$(VN_q)^{-1}\sum_\lambda C_\lambda v_\lambda^2$,
where $V$ is the cell volume, $N_q$ is the number of
$\bq$ points, and $v_\lambda^2$ is the squared group
velocity averaged over the three directions.
This factor and the $Cv^2$-weighted lifetime are
the two factors of $\kP$ and are reported relative
to perfect Si.
Their product gives the relative $\kP$, up to rounding.
The last column gives the largest atom-resolved
scattering contribution $N\Gamma_i$ in each cell,
relative to perfect Si.
}
\label{tab:defectcells}
\begin{ruledtabular}
\begin{tabular}{llcccccc}
Cell & Defects & Concentration & $\kappa$ & $\kP$
& $\sum Cv^2$ & Lifetime & Largest $N\Gamma_i$\\
& & (\%)
& \multicolumn{2}{c}{(W\,m$^{-1}$K$^{-1}$)}
& \multicolumn{3}{c}{(relative to Si)}\\
\hline
64 sites  & none          & 0   & 124.2 & 124.0 & 1    & 1    & 1\\
64 sites  & one vacancy   & 1.6 & 8.2   & 6.1   & 0.22 & 0.22 & 10.8\\
64 sites  & divacancy     & 3.1 & 6.0   & 4.7   & 0.21 & 0.18 & 9.8\\
128 sites & two vacancies & 1.6 & 6.5   & 4.3   & 0.18 & 0.19 & 12.8\\
64 sites  & one Ge        & 1.6 & 58.1  & 53.6  & 0.29 & 1.47 & 1.35\\
64 sites  & Ge pair       & 3.1 & 48.3  & 45.8  & 0.28 & 1.34 & 1.62\\
128 sites & two Ge        & 1.6 & 52.6  & 48.3  & 0.27 & 1.44 & 1.50\\
\end{tabular}
\end{ruledtabular}
\end{table*}

\begin{figure}[t]
\includegraphics[width=\linewidth]{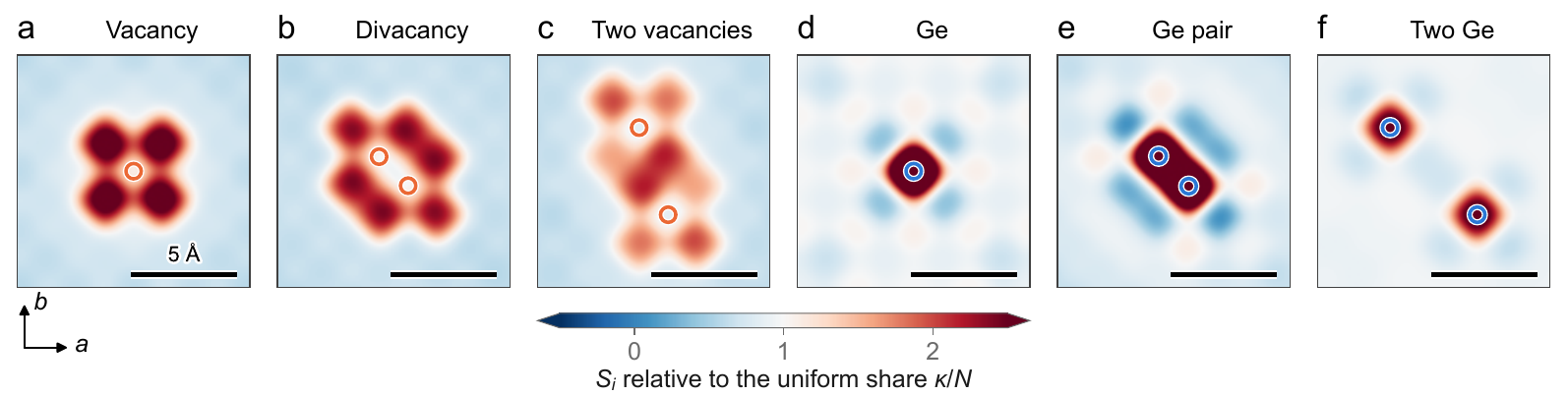}
\caption{$S_i$ relative to the uniform share $\kappa/N$ for the defects of Fig.~\ref{fig:defect}, projected along
$c$ and smoothed with a 0.7~\AA\ Gaussian. a. vacancy, b. divacancy (64-site cells), c. two vacancies (128-site cell),
d. Ge, e. Ge pair (64-site cells), f. two Ge atoms (128-site cell). Orange and blue circles mark vacancies and Ge
atoms.}
\label{fig:si_pairs}
\end{figure}

\textit{Sensitivity $S_i$.} Figure~\ref{fig:si_pairs} shows $S_i=S_i^{\mathrm P}+S_i^{\mathrm C}$ relative to the
uniform share $\kappa/N$. Here $S_i^{\mathrm P}$ is the sensitivity of Appendix~\ref{si:atom}, and $S_i^{\mathrm C}$
allocates $\kC$ to the atoms with the same weights $\Gamma_{\lambda,i}/\Gamma_\lambda$, so that $\sum_iS_i=\kappa$.
Around the vacancy, $S_i$ is concentrated on the four neighbours, which carry 43\% of the summed $S_i$ although they
are 6\% of the atoms, and no atom nearby has a negative value. Around Ge, $S_i$ is 12.8 times the uniform share on
the Ge atom and negative on its four neighbours ($-0.72$ each). Together the five atoms carry 15.5\% of the summed
$S_i$. $S_i$ can be negative because $\mathrm{Re}\,[\Phi^{(i)*}\Phi]$ contains the cross terms between atoms. For the
modes below 3~THz the four neighbours of the vacancy account for 92--143\% of the linewidth, with the remainder
spread as small negative terms over the other atoms, whereas the Ge atom takes 25--36\% and its four neighbours
$-6$ to $-20$\% of a linewidth that is 1.2--1.5 times that of Si. A possible reading is that the contributions of
the two atoms of a bond to the vertex of a long-wavelength phonon nearly cancel. A Ge atom keeps its four bonds, so
the change at the Ge end is largely cancelled at the other end, whereas at a vacancy the bond is missing.

\begin{figure}[t]
\includegraphics[width=0.8\linewidth]{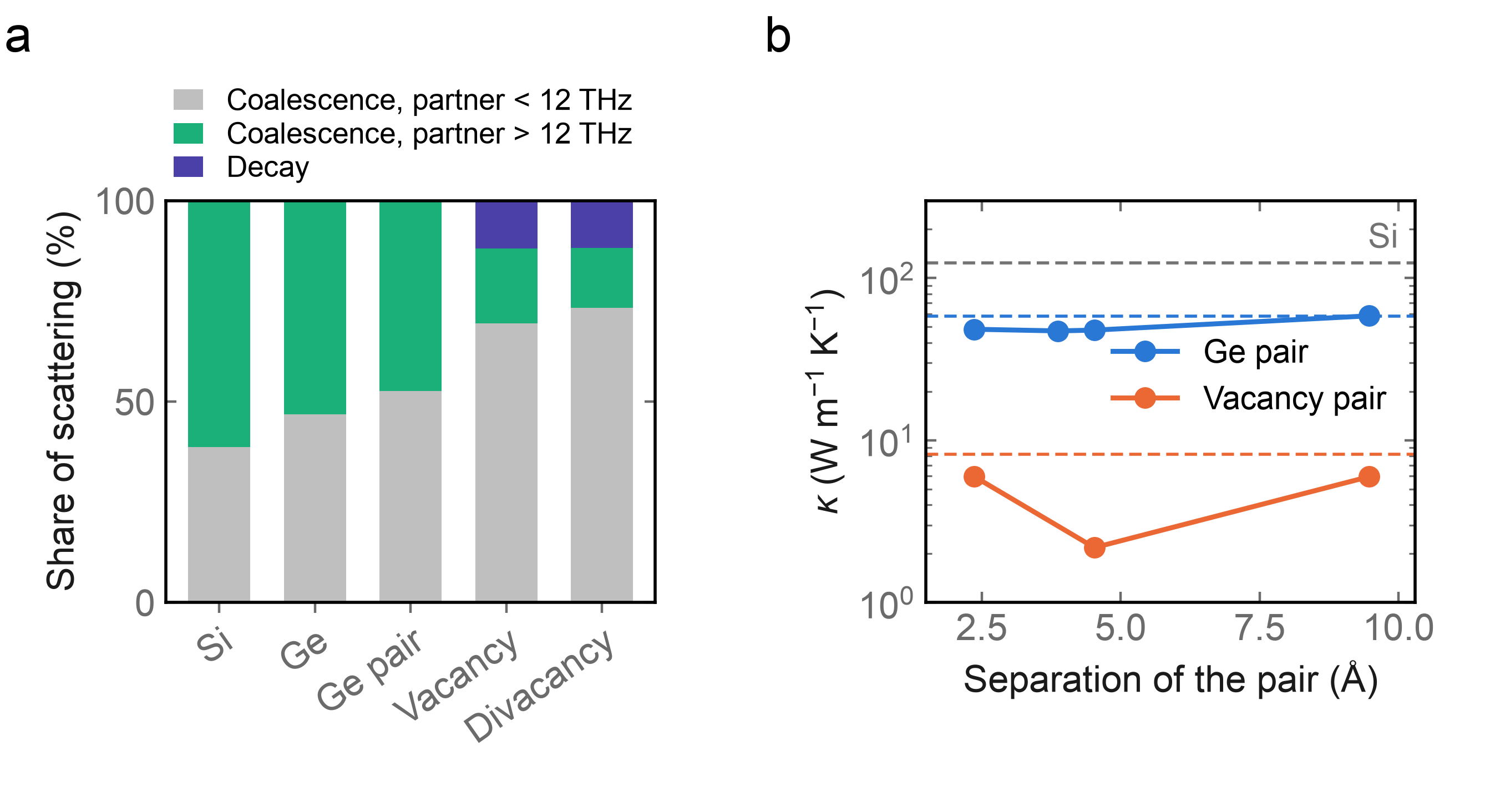}
\caption{a. Scattering of the heat carriers below 4~THz at 300~K, weighted by the mode contribution to $\kP$ and split into coalescence with a partner below 12~THz, coalescence with a partner above 12~THz, and decay. b. Total conductivity against the separation of two defects in the 64-site cell. Dashed lines give the cell with one defect of the same kind (blue Ge, orange vacancy) and perfect Si (grey).}
\label{fig:si_channels}
\end{figure}

\textit{Scattering channels.} Figure~\ref{fig:si_channels}a shows the scattering of the carriers below 4~THz, which hold 60\% of $\kP$ in Si, 76\% with Ge and 41\% with the vacancy. In Si most of it is coalescence with a partner above 12~THz, and decay is absent. Ge leaves this composition nearly unchanged. With vacancies the share of these processes falls to 15--19\%, and 12\% of the scattering is decay.

\textit{Defect pairs.} The arrangement of the defects matters at a fixed concentration (Fig.~\ref{fig:si_channels}b). These additional 64-site cells are not listed in Table~\ref{tab:defectcells}. Two Ge atoms 2.37 to 4.53~\AA\ apart give 47 to 48~W\,m$^{-1}$K$^{-1}$, and 58~W\,m$^{-1}$K$^{-1}$ when they are 9.47~\AA\ apart, while the scattering per Ge atom is the same. The difference comes from $\sum Cv^2$. Two vacancies 3.87~\AA\ apart are not stable. Without symmetry constraints the shared neighbour moves into one of them and they form a divacancy, 1.9~eV lower in energy than two separated vacancies. The divacancy and the pair 9.47~\AA\ apart give the same $\kappa$. A pair 4.53~\AA\ apart has a soft local mode (1.1~THz at $\Gamma$) and a lower $\kappa$ of 2.2~W\,m$^{-1}$K$^{-1}$, and in this cell only 9\% of $\kP$ is carried below 4~THz. A close pair also makes $\kappa$ anisotropic, with the largest principal value 6--10\% above the smallest for the Ge pairs and 38\% for the divacancy.

\section{Conductivities of the interface structures}\label{si:iface}

Table~\ref{tab:iface} lists the conductivities of the two interface structures of Fig.~\ref{fig:iface} by direction and temperature. Bulk diamond with the same potential, on a $19^3$ mesh at 300~K with natural isotopes, has 1755~W\,m$^{-1}$K$^{-1}$ in the RTA and 2241~W\,m$^{-1}$K$^{-1}$ from the direct solution of the linearised BTE.

\begin{table}[t]
\caption{Conductivities of the Si/diamond (001) superlattice (312 atoms, $6\times6\times2$ mesh) and of the diamond (001) $\Sigma5$ twist boundary (160 atoms, $8\times8\times4$ mesh) in W\,m$^{-1}$K$^{-1}$ (RTA and Wigner coherences, natural isotopes). ``In-plane'' is parallel to the interfaces and ``across'' is along $c$. The last column is the share of $\kC$ in $\kappa=\kP+\kC$.}
\label{tab:iface}
\begin{ruledtabular}
\begin{tabular}{llccccc}
structure & direction & $T$ (K) & $\kP$ & $\kC$ & $\kappa$ & $\kC/\kappa$ (\%)\\
\hline
Si/diamond & in-plane & 300 & 10.83 & 2.71 & 13.54 & 20\\
 & & 500 & 7.89 & 4.24 & 12.13 & 35\\
 & & 800 & 5.43 & 4.83 & 10.26 & 47\\
 & across & 300 & 0.162 & 0.234 & 0.396 & 59\\
 & & 500 & 0.095 & 0.380 & 0.475 & 80\\
 & & 800 & 0.058 & 0.535 & 0.593 & 90\\
\hline
Diamond boundary & in-plane & 300 & 172.1 & 0.89 & 173.0 & 0.5\\
 & & 500 & 70.3 & 2.09 & 72.4 & 2.9\\
 & & 800 & 37.9 & 3.18 & 41.1 & 7.7\\
 & across & 300 & 73.6 & 1.19 & 74.8 & 1.6\\
 & & 500 & 28.5 & 2.43 & 30.9 & 7.9\\
 & & 800 & 14.9 & 3.40 & 18.3 & 18.5\\
\end{tabular}
\end{ruledtabular}
\end{table}

\newpage
\bibliography{refs}

\end{document}

%% file: fig_overview.tex
\begin{figure}[t]
  \centering
  \begin{tikzpicture}[
    font=\PhonoFigureFont\footnotesize,
    execute at begin node={
      \setlength{\baselineskip}{1.2em}
      \hyphenpenalty=10000
      \exhyphenpenalty=10000
    },
    node distance=4mm and 5mm,
    box/.style={
      draw=black!55,
      rounded corners=2pt,
      fill=black!4,
      align=center,
      inner sep=4pt,
      minimum height=14mm
    },
    core/.style={
      draw=orange!80!black,
      very thick,
      rounded corners=2pt,
      fill=orange!10,
      align=center,
      inner sep=4pt,
      minimum height=17mm,
      text width=31mm
    },
    aux/.style={
      draw=blue!50!black,
      rounded corners=2pt,
      fill=blue!6,
      align=center,
      inner sep=4pt,
      minimum height=14mm
    },
    arr/.style={
      -{Stealth[length=2.2mm]},
      thick,
      black!70
    },
  ]

  % ===== Core linewidth calculations =====
  % Independent interaction and integration-weight paths

  \node[core] (k1) {
    \textbf{\ding{172} Fourier transform}\\[2pt]
    Third-order force constants\\
    to reciprocal space,\\
    batched contractions
  };

  \node[core, right=of k1] (k2) {
    \textbf{\ding{173} Mode projection}\\[2pt]
    Batched matrix products\\
    $\to|\Phi_{\lambda\lambda'\lambda''}|^2$
  };

  \node[
    aux,
    text width=27mm,
    minimum height=17mm,
    right=of k2
  ] (k4) {
    \textbf{Combine}\\[2pt]
    Weighted band sums,\\
    occupations for all $T$\\
    $\to\Gamma_\lambda(T)$
  };

  \node[core, right=of k4] (k3) {
    \textbf{\ding{174} Tetrahedron weights}\\[2pt]
    Branch-free regions,\\
    sorting,\\
    pruning of zero weights
  };

  % Interaction-strength path
  \draw[arr] (k1) -- (k2);
  \draw[arr] (k2) -- (k4);

  % Independent integration-weight path
  \draw[arr] (k3) -- (k4);

  % ===== Kernel background =====
  \coordinate (ktop) at ([yshift=7mm]k1.north);

  \begin{scope}[on background layer]
    \node[
      draw=orange!80!black,
      dashed,
      rounded corners=4pt,
      fill=orange!2,
      fit=(k1)(k2)(k3)(k4)(ktop),
      inner xsep=3mm,
      inner ysep=3mm
    ] (kernel) {};
  \end{scope}

  \node[
    anchor=north west,
    font=\PhonoFigureFont\footnotesize\bfseries,
    text=orange!45!black,
    inner sep=1pt
  ] at ([xshift=3mm,yshift=-2.5mm]kernel.north west) {
    Three-phonon linewidth (GPU):
    batches of triplets for each irreducible $\bq$
  };

  % ===== Top row: input and setup =====

  \node[
    box,
    text width=36mm,
    above=9mm of kernel.north -| k1,
    anchor=south
  ] (in) {
    \textbf{Input}\\
    $\Phi^{(2)},\Phi^{(3)}$ from DFT or MLIP,\\
    $\bq$ mesh,\\
    Born charges (optional)
  };

  \node[
    box,
    text width=36mm,
    at=(in -| k2),
    xshift=9mm
  ] (host) {
    \textbf{Setup, once (CPU)}\\
    Grids, irreducible $\bq$, triplets,\\
    tetrahedron tables
  };

  \node[
    aux,
    text width=36mm,
    at=(in -| k3)
  ] (harm) {
    \textbf{Harmonic phonons (GPU)}\\
    $\omega_\lambda$, eigenvectors,\\
    velocity operators
  };

  \draw[arr] (in) -- (host);
  \draw[arr] (host) -- (harm);
  \draw[arr]
    (harm.south) --
    (kernel.north -| harm.south);

  % ===== Bottom row: transport and analysis =====

  \node[
    aux,
    text width=31mm,
    below=9mm of kernel.south -| k1,
    anchor=north
  ] (iso) {
    \textbf{Isotope scattering (GPU)}\\
    Same tetrahedron weights\\
    $\to\Gamma^{\mathrm{iso}}_\lambda$
  };

  \node[
    aux,
    text width=31mm,
    at=(iso -| k2)
  ] (wig) {
    \textbf{Conductivity (GPU)}\\
    RTA and Wigner term\\
    $\kappa=\kP+\kC$
  };

  \node[
    box,
    text width=27mm,
    at=(iso -| k4)
  ] (out) {
    \textbf{Output}\\
    $\kappa$ and $\Gamma_\lambda(T)$\\
    in phono3py format
  };

  \node[
    box,
    text width=31mm,
    fill=green!6,
    draw=green!40!black,
    at=(iso -| k3)
  ] (atom) {
    \textbf{Atom-resolved analysis}\\
    $\Gamma_{\lambda,i}$, $\Gamma_i$, and $S_i$\\
    (separate run)
  };

  \draw[arr] (kernel.south -| iso) -- (iso);
  \draw[arr] (kernel.south -| wig) -- (wig);
  \draw[arr] (iso) -- (wig);
  \draw[arr] (wig) -- (out);

  % Optional analysis reusing the linewidth workflow
  \draw[arr, dashed]
    (kernel.south -| atom) -- (atom);

  \end{tikzpicture}

\caption{
Computational workflow of Phono3GPU.
The CPU prepares the grids, symmetry operations, and phonon triplets using the phono3py Python API. The GPU independently evaluates three-phonon interaction strengths through batched contractions and tetrahedron integration weights through branch-free operations and pruning. These quantities are combined to obtain phonon linewidths and thermal conductivity within the RTA and Wigner formulation. The atom-resolved analysis reuses the same computational workflow in a separate run. Further details are provided in Appendix~\ref{app:method}.
}
  \label{fig:overview}
\end{figure}